\documentclass[aps,prb,twocolumn,superscriptaddress,
groupedaddress,10pt]{revtex4}
\usepackage{amsmath}
\usepackage{amssymb}
\usepackage{graphicx}
\usepackage{epsfig}
\usepackage{dcolumn}
\usepackage{bm}
\usepackage{blindtext}
\usepackage{natbib}
\usepackage{siunitx} 
\usepackage{multirow}
\usepackage{bbm}

\usepackage[titletoc]{appendix}

\usepackage{simplewick}

\usepackage{bbm}
\makeatletter
\newcommand{\mathleft}{\@fleqntrue\@mathmargin0pt}
\newcommand{\mathcenter}{\@fleqnfalse}
\makeatother

\usepackage[usenames,dvipsnames]{xcolor}
\usepackage{amsthm}
\usepackage{booktabs}
\usepackage{hyperref}
\usepackage{tikz}
\usetikzlibrary{calc}
\usepackage{mathtools}

\def\be{\begin{equation}} \def\ee{\end{equation}}
\def\bea{\begin{eqnarray}} \def\eea{\end{eqnarray}}

\def\nn{\nonumber}

\begin{document}
\title{
Emergent SU(2)$_1$ conformal symmetry in spin-1/2 Kitaev-Gamma chain with a Dzyaloshinskii-Moriya interaction
}

\author{Wang Yang}
\affiliation{School of Physics, Nankai University, Tianjin, 300071, China}

\author{Alberto Nocera}
\affiliation{Department of Physics and Astronomy and Stewart Blusson Quantum Matter Institute, 
University of British Columbia, Vancouver, B.C., Canada, V6T 1Z1}

\author{Chao Xu}
\affiliation{Kavli Institute for Theoretical Sciences, University of Chinese Academy of Sciences, Beijing 100190, China}
\affiliation{Institute for Advanced Study, Tsinghua University, Beijing 100084, China}

\author{Shicheng Ma}
\affiliation{School of Physics, Nankai University, Tianjin, 300071, China}

\author{Arnab Adhikary}
\affiliation{Department of Physics and Astronomy and Stewart Blusson Quantum Matter Institute, 
University of British Columbia, Vancouver, B.C., Canada, V6T 1Z1}


\author{Ian Affleck}
\affiliation{Department of Physics and Astronomy and Stewart Blusson Quantum Matter Institute, 
University of British Columbia, Vancouver, B.C., Canada, V6T 1Z1}

\begin{abstract}
We study the one-dimensional spin-1/2 Kitaev-Gamma model with a bond-dependent Dzyaloshinskii-Moriya (DM) interaction, which can be induced by an electric field applied in the third direction where the first and second directions refer to the two bond directions in the model. 
By a combination of field theory and symmetry analysis, extended gapless phases with an emergent SU(2)$_1$ conformal symmetry are  found in the phase diagram of the spin-1/2 Kitaev-Gamma-DM chain. 
The analytic predictions are in good agreements with numerical results obtained from density matrix renormalization group simulations.  

\end{abstract}
\maketitle

\section{Introduction}

As potential realizations of the Kitaev spin-1/2 model on the two-dimensional (2D) honeycomb lattice which is useful for topological quantum computation \cite{Kitaev2006,Nayak2008},
Kitaev materials (including A$_2$IrO$_3$ (A$=$Li, Na), $\alpha$-RuCl$_3$, etc.)  have attracted intensive research attentions in the past decade \cite{Jackeli2009,Chaloupka2010,Singh2010,Price2012,Singh2012,Plumb2014,Kim2015,Winter2016,Baek2017,Leahy2017,Sears2017,Wolter2017,Zheng2017,Rousochatzakis2017,Kasahara2018,Rau2014,Ran2017,Wang2017,Catuneanu2018,Gohlke2018,Liu2011,Chaloupka2013,Johnson2015,Motome2020}. 
Motivated by the theoretical difficulties in strongly correlated systems in 2D, there has been a surge of research interests in studying 1D generalized Kitaev spin models \cite{Sela2014,Agrapidis2018,Agrapidis2019,Catuneanu2019,Yang2020,Yang2020a,Yang2020b,Yang2021b,Yang2022,Yang2022b,Yang2020c,Yang2020d,Luo2021,Luo2021b,You2020,Sorensen2021}, which can be constructed by selecting one or several rows out of the honeycomb lattice. 
The hope is that such 1D studies can help understand the 2D Kitaev physics\cite{Yang2022b,Yang2020d}.

Besides providing hints for 2D physics, the 1D Kitaev spin models are also intriguing on their own, containing exotic strongly correlated physics. 
For example, in the 1D spin-1/2 Kitaev-Gamma model, it has been found that about 67\% of the entire phase diagram is occupied by a gapless phase  whose low energy physics can be described by an emergent SU(2)$_1$ Wess-Zumino-Witten (WZW) model,
which is intimately related to the intricate discrete nonsymmorphic symmetry group of the system \cite{Yang2020}. 
In addition to the emergent SU(2)$_1$ phase, the spin-1/2 Kitaev-Gamma chain also contains two magnetically ordered phases with symmetry breaking patterns $O_h\rightarrow D_4$ \cite{Yang2020} and $O_h\rightarrow D_3$ \cite{Yang2021b}, as well as another gapless phase with central charge $c=1/2$ located in the vicinity of the antiferromagnetic (AFM) Kitaev point \cite{Luo2021} , where $O_h$ is the full octahedral group and $D_n$ is the dihedral group of order $2n$, showcasing rich strongly correlated physics in such systems. 
Recently, it has been established that CoNb$_2$O$_6$ is a one-dimensional Kitaev material with ferromagnetic (FM) Kitaev and AFM Gamma interactions \cite{Churchill2024}. 

On the other hand, nearest neighbor Dzyaloshinskii-Moriya (DM) interactions \cite{Dzyaloshinskii1957,Dzyaloshinskii1964,Moriya1960,Fert1980,Fert1990} are ubiquitous in quantum magnetic systems which break the spatial inversion symmetry and can lead to exotic magnetic textures such as skyrmions \cite{Heinze2011,Yu2010,Bogdanov1994,Rolber2006} and chiral magnetic orders \cite{Dzyaloshinskii1965,Bogdanov1989}.  
In Kitaev materials, the DM interactions can be conveniently induced and manipulated by external electric fields,
which have been studied under the context of  the electromagnetic controls of Kitaev materials \cite{Furuya2021,Chari2021}. 
It is noteworthy to mention that the DM interactions in Kitaev materials exhibit a nonuniform bond-dependent structure.

\begin{figure}[ht]
\begin{center}
\includegraphics[width=8cm]{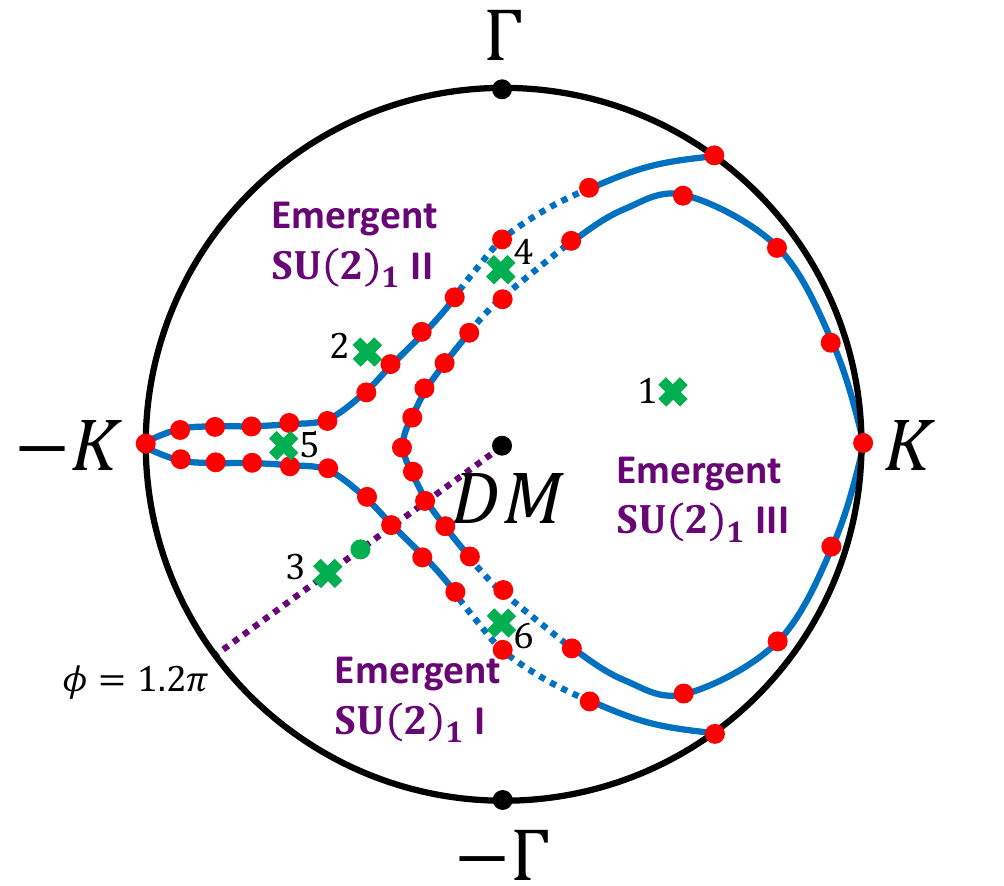}
\caption{
Emergent $\textrm{SU(2)}_1$ phases in the phase diagram of the spin-1/2 Kitaev-Gamma chain with a DM interaction. 
In terms of the parametrization in Eq.~(\ref{eq:parametrize_KGDM}), the points $DM$, $K$, $-K$, $\Gamma$, $-\Gamma$ in the figure are parametrized by $\theta=0$, $(\theta=\pi/2,\phi=0)$, $(\theta=\pi/2,\phi=\pi)$, $(\theta=\pi/2,\phi=\pi/2)$, and $(\theta=\pi/2,\phi=3\pi/2)$, respectively. 
The red solid circles are phase transition points determined by calculating single-copy entanglement. 
Green crosses are points 
where central charges are calculated in Sec. \ref{subsec:phase_transition},
in order to further confirm the ranges of the emergent $\textrm{SU(2)}_1$ phases.
The purple dashed line represents the line of $\phi=1.2\pi$, 
which is the parameter chosen in Sec. \ref{subsec:SU2_1} in DMRG numerical simulations. 
The green solid circle corresponds to $(K=\cos(1.2\pi),\Gamma=\sin(1.2\pi),D_M=1.0)$ up to normalization. 
} 
\label{fig:phase_KGDM}
\end{center}
\end{figure}

In this work, we study the effects of a DM interaction in the 1D spin-1/2 Kitaev-Gamma model, which can be created by applying an electric field along the third direction (where the first and second directions refer to the two bond directions in the 1D model).
The obtained phase diagram is shown in Fig. \ref{fig:phase_KGDM},
in which large regions of emergent SU$(2)_1$ conformal symmetries at low energies have been found,
denoted as ``Emergent SU$(2)_1$ I, II, III" in the figure.  
Since in real Kitaev materials, the Kitaev and Gamma interactions usually dominate over other interactions such as the Heisenberg and $\Gamma^\prime$ terms,
our work  provides a perturbative starting point for the  analysis of more complicated interactions as well as an extrapolation  to 2D in the presence of a DM interaction. 

We start from a pure Kitaev spin-1/2 chain with a DM interaction. 
The first important result of our work is that there is a two-site periodic unitary transformation $V_2$, which maps the 1D Kitaev-DM model to the Kitaev-Gamma model and vice versa. 
Using the known phase diagram of the 1D spin-1/2 Kitaev-Gamma model,
the effects of the DM interaction on the spin-1/2 Kitaev chain is thereby fully understood.

We then proceed to studying the 1D Kitaev-Gamma-DM model.
The strategy is to take the emergent SU(2)$_1$ phase in the Kitaev-Gamma chain as the unperturbed system, and treat the DM interaction as a perturbation. 
We find that although there is a dimension $3/2$ operator in the low energy conformal field theory (CFT), such operator is a total derivative and vanishes if periodic boundary condition is imposed. 
Therefore, even with a nonzero DM interaction, the system remains to have an emergent SU(2)$_1$ conformal symmetry in an extended region in the phase diagram. 
The analytical  predictions are in good agreements with our  density matrix renormalization group (DMRG) numerical simulations. 

Two observations are in order. 
First, the aforementioned unitary transformation $V_2$ is a duality transformation which maps the Kitaev-Gamma-DM model to itself with a different set of parameters,
thereby enlarging the region occupied by the emergent SU(2)$_1$ phase in the phase diagram. 
Second, it is interesting to compare the symmetry groups of the Kitaev-Gamma and Kitaev-Gamma-DM models. 
The symmetry analysis is facilitated by a six-sublattice rotation $U_6$ \cite{Yang2020} which hold for both models.
In the $U_6$ frame, it has been shown in Ref. \onlinecite{Yang2020} that the symmetry group $G$ of the Kitaev-Gamma model satisfies $G/\mathopen{<}T_{3a}\mathclose{>}\cong O_h$ where $T_{na}$ represents the translation operator by $n$ sites.
On the other hand, the symmetry group $G_z$ of the Kitaev-Gamma-DM model is found to satisfy $G_z/\mathopen{<}T_{6a}\mathclose{>}\cong O_h$.
Hence, the symmetry group in the Kitaev-Gamma-DM model is ``halved" compared with the Kitaev-Gamma model.
Notice that the unit cell of the spin-1/2 Kitaev-Gamma-DM model contains an even number of sites, which naively corresponds to an integer spin. 
Hence it is rather an unexpected result that such ``integer" spin system displays an emergent gapless phase at low energies,
which essentially originates from the intricate nonsymmorphic group structure. 

The rest of the paper is organized as follows.
Sec. \ref{sec:model} introduces the model Hamiltonian and discusses the unitarily equivalent relations in the model. 
In addition, two exactly solvable points are found  with equal strength of Gamma and DM couplings, which have an exponentially large ground state degeneracy. 
In Sec. \ref{sec:KGDM_SU2_1}, the emergent SU(2)$_1$ conformal symmetry is obtained by treating the DM interaction as a perturbation on the known SU(2)$_1$ phase in the Kitaev-Gamma model.
Sec. \ref{sec:low_theory_DM_z} presents a symmetry analysis, which proves that the analysis in Sec. \ref{sec:KGDM_SU2_1} has exhausted all possible relevant and marginal operators in the low energy field theory. 
Sec. \ref{sec:numerics} shows the numerical evidence for the emergent SU(2)$_1$ conformal symmetry
and numerically determines the phase boundaries of the emergent SU(2)$_1$ phases. 
Finally in Sec. \ref{sec:summary}, we summarize the main results of this work.

\section{The model}
\label{sec:model}

\subsection{Model Hamiltonian}

\begin{figure}[ht]
\begin{center}
\includegraphics[width=8cm]{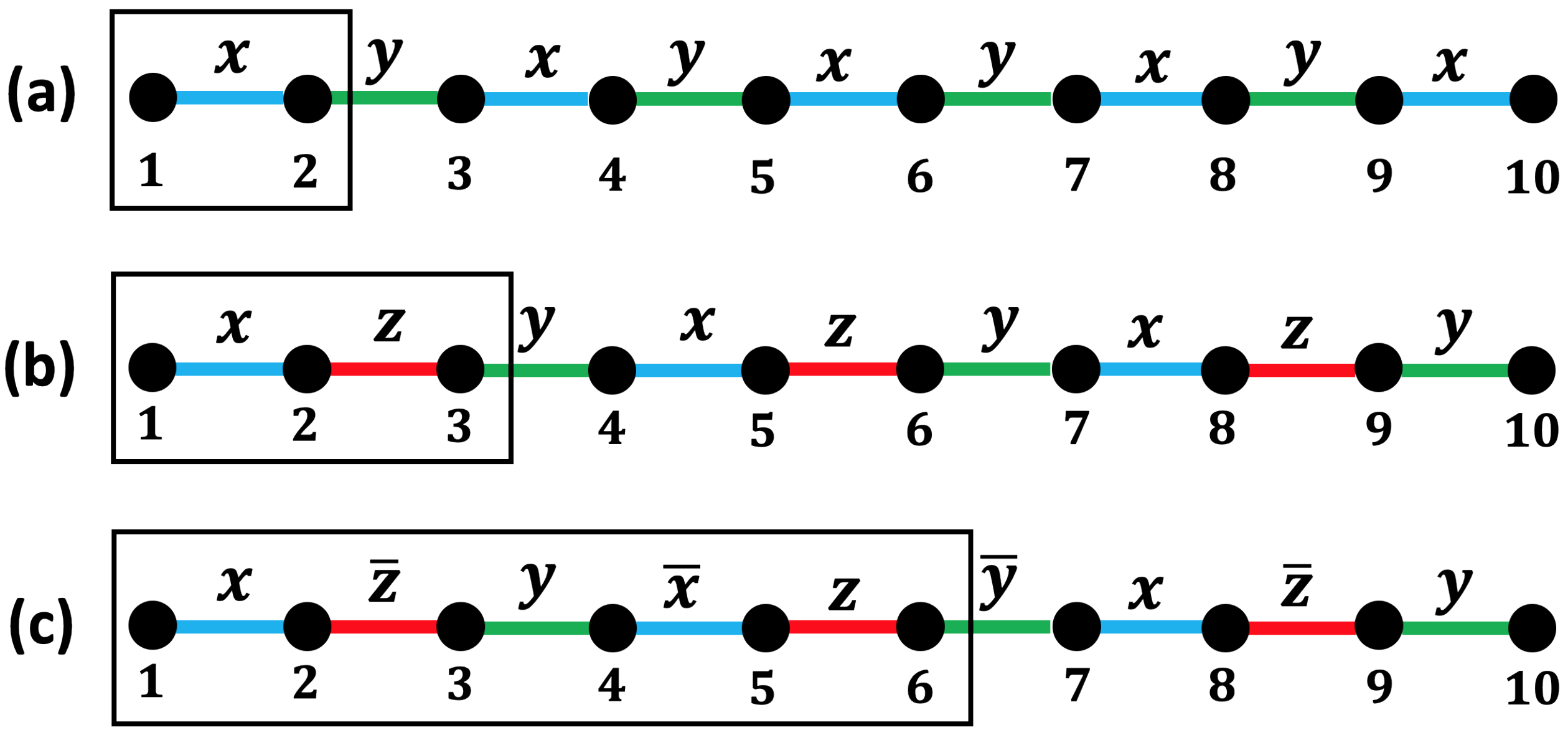}
\caption{Bond patterns of the Kitaev-Gamma chain (a)  without sublattice rotation,
(b) after the six-sublattice rotation,
(c) with a nonzero DM interaction after six-sublattice rotation.
The black squares represent the unit cells. 
} \label{fig:bond_pattern}
\end{center}
\end{figure}

The Hamiltonian of the spin-1/2 Kitaev-Gamma chain is defined as 
\begin{eqnarray}
H=\sum_{\langle ij \rangle\in\gamma\,\text{bond}}\big[KS_i^\gamma S_j^\gamma+ \Gamma (S_i^\alpha S_j^\beta+S_i^\beta S_j^\alpha)\big],
\label{eq:Ham_KG}
\end{eqnarray}
in which $(\alpha,\beta,\gamma)$ form a right-handed coordinate system and
the pattern for the bond $\gamma$ is shown in Fig. \ref{fig:bond_pattern} (a).
A useful unitary transformation is the  six-sublattice rotation $U_6$, defined as 
\begin{eqnarray}
\text{Sublattice $1$}: & (x,y,z) & \rightarrow (x^{\prime},y^{\prime},z^{\prime}),\nn\\ 
\text{Sublattice $2$}: & (x,y,z) & \rightarrow (-x^{\prime},-z^{\prime},-y^{\prime}),\nn\\
\text{Sublattice $3$}: & (x,y,z) & \rightarrow (y^{\prime},z^{\prime},x^{\prime}),\nn\\
\text{Sublattice $4$}: & (x,y,z) & \rightarrow (-y^{\prime},-x^{\prime},-z^{\prime}),\nn\\
\text{Sublattice $5$}: & (x,y,z) & \rightarrow (z^{\prime},x^{\prime},y^{\prime}),\nn\\
\text{Sublattice $6$}: & (x,y,z) & \rightarrow (-z^{\prime},-y^{\prime},-x^{\prime}),
\label{eq:6rotation}
\end{eqnarray}
in which ``Sublattice $i$" ($1\leq i \leq 6$) represents all the sites $i+6n$ ($n\in \mathbb{Z}$) in the chain, and we have abbreviated $S^\alpha$ ($S^{\prime \alpha}$) as $\alpha$ ($\alpha^\prime$) for short ($\alpha=x,y,z$).
In the six-sublattice rotated frame, the Hamiltonian in Eq. (\ref{eq:Ham_KG}) becomes 
\bea
H^\prime=\sum_{\langle ij \rangle\in\gamma\,\text{bond}}\big[-KS_i^\gamma S_j^\gamma - \Gamma (S_i^\alpha S_j^\alpha+S_i^\beta S_j^\beta)\big],
\label{eq:Ham_rot0}
\eea
in which the pattern for the bond $\gamma$ is shown in Fig. \ref{fig:bond_pattern} (b), having a three-site periodicity. 

In Kitaev materials, nearest neighbor DM interactions can be induced by applying electric fields.
Let $\gamma$ be the bond connecting nearest neighboring sites $i$ and $j$, and consider the right-handed coordinate system $(\gamma,\alpha,\beta)$.
An electric field is called in-plane if its direction is within the $\alpha\beta$-plane, and out-of-plane if it is parallel with the $\gamma$-axis. 
As discussed in Ref. \onlinecite{Furuya2021}, 
an in-plane electric field induces a DM interaction of the form
\bea
D_{M}(S_i^\alpha S_{i+1}^\beta-S_i^\beta S_{i+1}^\alpha),
\label{eq:D_Fz}
\eea
in which $D_M$ is given by \cite{Furuya2021}
\begin{flalign}
&D_M=\nonumber\\
&-\frac{4IJ_F}{t}\frac{(E_\alpha+E_\beta)(U_d-U_p+\Delta_{dp})+(E_\alpha-E_\beta)J_H}{2(U_d-U_p)+J_H},
\end{flalign}
where $E_\alpha$ and $E_\beta$ are the components of the electric  fields along $\alpha$- and $\beta$-directions;
$t$ is the hopping integral between $d$- and $p$-orbitals;
$U_d$ and $U_p$ are the strengths of on-site Hubbard interaction on $d$- and $p$-orbitals;
$J_H$ is the Hund's coupling at $p$-orbitals;
$\Delta_{dp}$ is the orbital energy difference between $d$- and $p$-orbitals;
$I$ is the matrix element of the electric dipole moment between $d$- and $p$-orbitals;
and $J_F$ is given by 
\begin{flalign}
J_F=-\frac{8}{3}t^4 \frac{1}{2(U_d-U_p+\Delta_{dp})^2 [(U_d-U_p+\Delta_{dp})-J_H]}.
\end{flalign}

In this work, we consider the type of DM interaction which can be induced by an electric field along the $z$-direction.
As can be seen from  Fig. \ref{fig:bond_pattern} (a), an electric field along z-direction corresponds to an in-plane electric field for every bond in the chain.
In the small $J_H$ limit, namely, $J_H\ll U_d-U_p$, at a first approximation, 
the value of the DM interaction remains the same at every bond, given by 
\begin{eqnarray}
D_M=
-\frac{4IJ_F}{t}\frac{E_z(U_d-U_p+\Delta_{dp})}{2(U_d-U_p)},
\label{eq:D_M_z}
\end{eqnarray} 
in which $E_z$ is the strength of the electric field.  
We note that the effect of a small $J_H$ can be analytically treated as a perturbation on the to-be-obtained critical theory for the $J_H=0$ case.
Adding Eq. (\ref{eq:D_Fz}) to Eq. (\ref{eq:Ham_KG}),
the Hamiltonian becomes
\begin{eqnarray}
H_z=\sum_{\langle ij \rangle\in\gamma\,\text{bond}}\big[KS_i^\gamma S_j^\gamma+ \Gamma_1 S_i^\alpha S_j^\beta+\Gamma_2 S_i^\beta S_j^\alpha\big],
\label{eq:Ham_KG_z}
\end{eqnarray}
in which 
\bea
\Gamma_1=\Gamma+D_M,~\Gamma_2=\Gamma-D_M;
\eea
the bond pattern for $\gamma$ is shown in Fig. \ref{fig:bond_pattern} (a);
and  $(\gamma\alpha\beta)$ form a right-handed coordinate system.
 More explicitly, the form of the Hamiltonian within a two-site unit cell of sites $\{1,2\}$ is given by
\bea
H_{z,12} &=&  K S^x_1 S^x_2 + \Gamma_1 S^y_1 S^z_2+\Gamma_2 S^z_1 S^y_2,\nn \\
H_{z,23} &=& K S^y_2 S^y_3 + \Gamma_1 S^z_2 S^x_3+\Gamma_2 S^x_2 S^z_3.
\label{eq:H}
\eea




In the six-sublattice rotated frame, the Hamiltonian is
\bea
H^\prime_z=\sum_{\langle ij \rangle\in\gamma\,\text{bond}}\big[-KS_i^\gamma S_j^\gamma - \Gamma_1 S_i^\alpha S_j^\alpha- \Gamma_2 S_i^\beta S_j^\beta\big],
\label{eq:Ham_rot}
\eea
in which
$\gamma\in\{x,\bar{z},y,\bar{x},z,\bar{y}\}$ has a six-site periodicity as shown in Fig. \ref{fig:bond_pattern} (c);
$S_j^\gamma=S_j^{\bar{\gamma}}$;
 $(\gamma\alpha\beta)$ form a right-handed coordinate system for $\gamma\in\{x,y,z\}$, and they form a left-handed system when $\gamma\in\{\bar{x},\bar{y},\bar{z}\}$.
 More explicitly, the form of the Hamiltonian within a six-site unit cell of sites $\{1,2,3,4,5,6\}$ is
 \bea
H^{\prime}_{z,12} &=&  -K S^{ x}_1 S^{ x}_2-\Gamma_1 S^{ y}_1 S^{ y}_2-\Gamma_2 S^{ z}_1 S^{ z}_2, \nn \\
H^{\prime}_{z,23} &=& -K S^{z }_2 S^{ z}_3-\Gamma_1 S^{ y}_2 S^{ y}_3-\Gamma_2 S^{ x}_2 S^{ x}_3, \nn \\
H^{\prime}_{z,34} &=& -K S^{ y}_3 S^{ y}_4-\Gamma_1 S^{ z}_3 S^{ z}_4-\Gamma_2 S^{ x}_3 S^{ x}_4, \nn \\
H^{\prime}_{z,45} &=& -K S^{ x}_4 S^{ x}_5-\Gamma_1 S^{ z}_4 S^{ z}_5-\Gamma_2 S^{ y}_4 S^{ y}_5, \nn \\
H^{\prime}_{z,56} &=& -K S^{z }_5 S^{ z}_6-\Gamma_1 S^{ x}_5 S^{ x}_6-\Gamma_2 S^{ y}_5 S^{ y}_6, \nn \\
H^{\prime}_{z,67} &=& -K S^{ y}_6 S^{ y}_7-\Gamma_1 S^{ x}_6 S^{ x}_7-\Gamma_2 S^{ z}_6 S^{ z}_7.
\label{eq:Hprime}
\eea
From here on, we will stick to the six-sublattice rotated frame in this work unless otherwise stated.
 
Finally, a useful parametrization is 
\bea
D_M&=&\cos(\theta),\nn\\
K&=&\sin(\theta)\cos(\phi),\nn\\
\Gamma&=&\sin(\theta)\sin(\phi),
\label{eq:parametrize_KGDM}
\eea
in which $\theta\in(0,\pi)$, $\phi\in(0,2\pi)$.
Hence, the full parameter  space of the model is a unit sphere.


\subsection{Unitarily equivalent relations}


Since the DM interaction breaks inversion symmetry,
the first observation is that in the original frame  (i.e., without six-sublattice rotation $U_6$), 
a spatial inversion with respect to any bond center flips the sign of the DM term while leaving the Kitaev and Gamma couplings unchanged. 
Therefore, we have the following equivalence relation 
\bea
(K,\Gamma,D_M)\simeq (K,\Gamma,-D_M),
\label{eq:equiv0}
\eea
in which the symbol ``$\simeq$" is used to indicate that the Hamiltonians defined by the two set of parameters are related by a unitary transformation. 
Second,  both $\Gamma_1$ and $\Gamma_2$ change sign under $R(\hat{z},\pi)$ within the original frame.
Hence there is the equivalence
\bea
(K,\Gamma,D_M)\simeq (K,-\Gamma,-D_M).
\label{eq:equiv1}
\eea
Furthermore,  consider a two-sublattice rotation $V_2$  in the original frame which acts on odd sites as a spin rotation $R(\hat{z},\pi)$ but leaves all even sites unchanged. 
Straightforward calculations show that the Kitaev, Gamma and DM terms transform under $V_2$ as
\bea
KS_i^\gamma S_j^\gamma &\rightarrow& -KS_i^\gamma S_j^\gamma,\nn\\
\Gamma (S_i^\alpha S_j^\beta+S_i^\beta S_j^\alpha) &\rightarrow& -\Gamma (S_i^\alpha S_j^\beta-S_i^\beta S_j^\alpha),\nn\\
D_{M}(S_i^\alpha S_{i+1}^\beta-S_i^\beta S_{i+1}^\alpha) &\rightarrow& -D_M (S_i^\alpha S_{i+1}^\beta+S_i^\beta S_{i+1}^\alpha),\nn\\
\label{eq:equiv_V6_KGammaDM}
\eea
which hold for every bond. 
Hence, we also have the following equivalence
\bea
(K,\Gamma,D_M)\simeq (-K,-D_M,-\Gamma).
\label{eq:equivalence}
\eea

\begin{figure}[ht]
\begin{center}
\includegraphics[width=7cm]{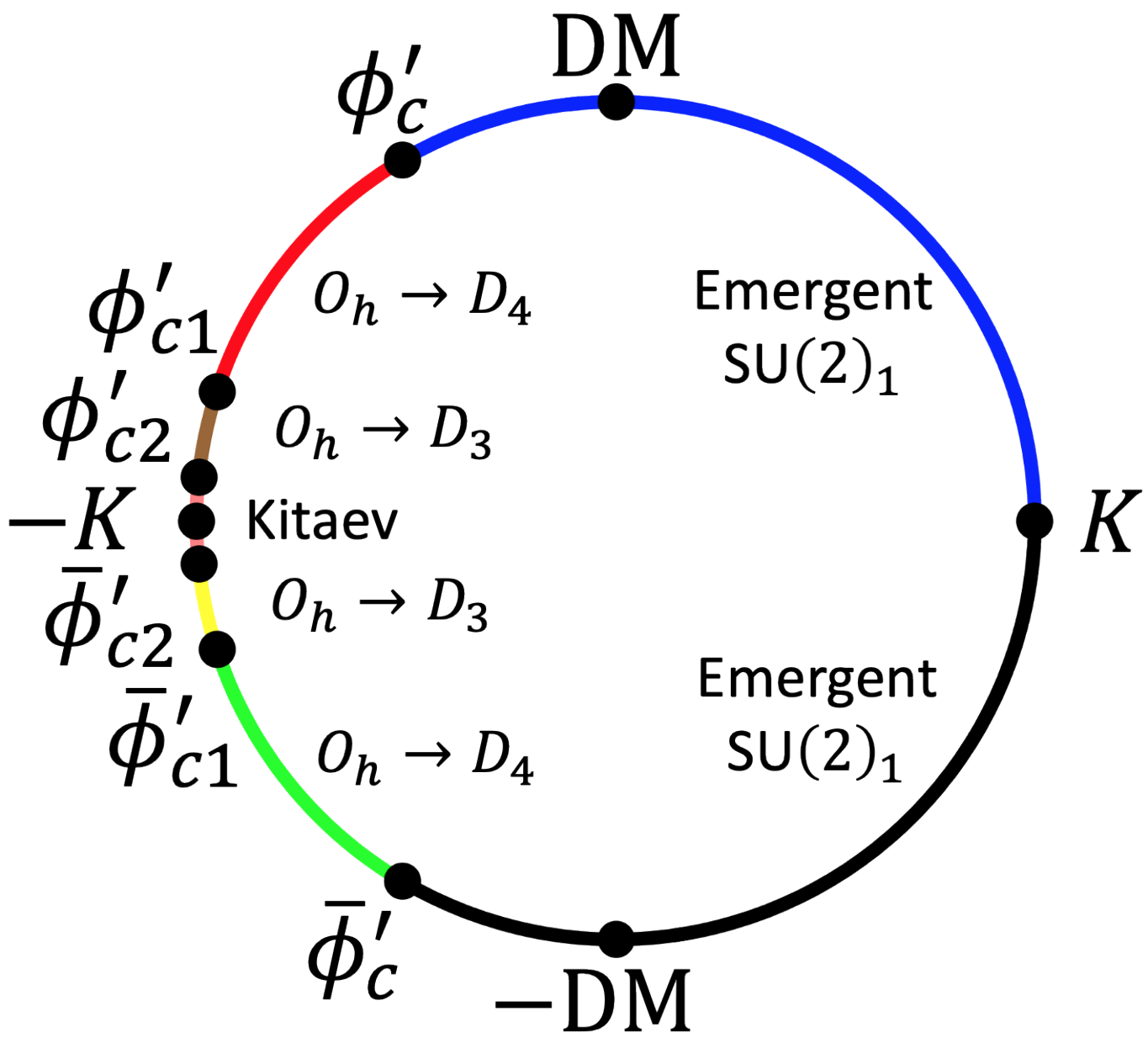}
\caption{Phase diagram of the spin-1/2 Kitaev chain with DM interaction, where $K=\cos(\phi^\prime)$, $D_M=\sin(\phi^\prime)$.
} \label{fig:phase_KDM}
\end{center}
\end{figure}

Using Eq. (\ref{eq:equivalence}), it can be seen that the Kitaev-DM model can be mapped to the  Kitaev-Gamma model.
On the other hand, the phase diagram of the spin-1/2 Kitaev-Gamma chain has been studied in detail in Ref. \onlinecite{Yang2020} and Ref. \onlinecite{Yang2021b}.
Therefore, we are able to directly obtain the phase diagram of the spin-1/2 Kitaev chain with a DM interaction as shown in Fig. \ref{fig:phase_KDM},
in which $K$ and $D_M$ are parametrized as
\bea
K=\cos(\phi^\prime),~D_M=\sin(\phi^\prime),
\eea
and the values of the phase transition points are
$\phi^\prime_c=0.67\pi$, $\phi^\prime_{c1}=0.9\pi$, $\phi^\prime_{c2}=0.967\pi$,
$\bar{\phi^\prime}_c=2\pi-\phi^\prime_c$, $\bar{\phi}^\prime_{c1}=2\pi-\phi^\prime_{c1}$, and $\bar{\phi}^\prime_{c2}=2\pi-\phi^\prime_{c2}$.
Since $D_M$  changes sign under $R(\hat{z},\pi)$ whereas $K$ remains the same,
it is enough to consider the region $\phi^\prime\in[0,\pi)$.
For $\phi^\prime\in(0,\phi_c)$, the system is in a gapless phase which has an emergent SU(2)$_1$ conformal symmetry at low energies;
for $\phi^\prime\in(\phi^\prime_c,\phi^\prime_{c1})$, the system is in an ordered phase with a spontaneous symmetry breaking pattern as $O_h\rightarrow D_4$;
for $\phi^\prime\in(\phi^\prime_{c1},\phi^\prime_{c2})$, the system is in another ordered phase with a spontaneous symmetry breaking pattern as $O_h\rightarrow D_3$;
for $\phi^\prime\in(\phi^\prime_{c2},\pi)$ named as ``Kitaev" in Fig. \ref{fig:phase_KDM}, there is numerical evidence that the central charge in this region is $c=1/2$ \cite{Luo2021}.

\subsection{Exactly solvable points at $D_M=\pm \Gamma$}
\label{subsec:solvable}

It is known that the Kitaev points are exactly solvable \cite{Brzezicki2007} regardless of the sign of the Kitaev coupling. 
The 1D pure Kitaev model has an exponentially large ground state degeneracy \cite{Brzezicki2007,You2008}, making it numerically very difficult to investigate the regions nearby in the phase diagram \cite{Yang2020}. 
In this section, we demonstrate that in the absence of the Kitaev coupling, the 1D Gamma-DM model is exactly solvable when $D_M=\pm \Gamma$, also having an exponentially large ground state degeneracy.

Since $\Gamma=D_M$ and $\Gamma=-D_M$ are related by inversion operation, it is enough to consider the $\Gamma=D_M$ case,
where the Hamiltonian is given by Eq. (\ref{eq:H}) by setting $K=0$, $\Gamma_1=-2\Gamma$, $\Gamma_2=0$.
Performing a two-sublattice unitary transformation $W_2$ which maps $(S^x_{2n},S^y_{2n},S^z_{2n})$ on even sites to $(-S^z_{2n},-S^y_{2n},-S^x_{2n})$ while leaving the spin operators on odd sites unchanged,
the Hamiltonian becomes
\bea
H_{e.s.}=\Gamma_1 \sum_{n}(S_{2n-1}^y S_{2n}^x+S_{2n}^xS_{2n+1}^x).
\label{eq:solvable}
\eea
It turns out that $H_{e.s.}$ is exactly solvable via the following Jordan-Wigner transformation,
\bea
S_j^x&=&\frac{1}{2}\gamma_j \Pi_{k<j}(i\gamma_k\gamma^\prime_k),\nn\\
S_j^y&=& \frac{1}{2}\gamma^\prime_j \Pi_{k<j}(i\gamma_k\gamma^\prime_k),\nn\\
S_j^z&=& -\frac{1}{2}i\gamma_j\gamma_j^\prime,
\eea
in which $\gamma_j$ and $\gamma_j^\prime$ are Majorana fermions satisfying 
\begin{flalign}
&(\gamma_j)^2=(\gamma_j^\prime)^2=1, \nn\\
&\{\gamma_j,\gamma_k\}=\{\gamma^\prime_j,\gamma^\prime_k\}=0,~j\neq k,\nn\\
&\{\gamma_j,\gamma^\prime_k\}=0,~\forall j,k.
\end{flalign}
These Majorana fermions can be mapped to a single-component spinless fermion by the following relations 
\bea
\gamma_j&=&c_j^\dagger+c_j,\nn\\
\gamma_j^\prime&=&-i(c_j^\dagger-c_j).
\eea

Using Majorana fermions $\gamma_j,\gamma_j^\prime$, the Hamiltonian $H_{e.s.}$ can be written as  
\bea
H_{e.s.}=\frac{1}{4}\Gamma_1\sum_n (i\gamma_{2n-1}\gamma_{2n}-i\gamma^\prime_{2n}\gamma_{2n+1}).
\label{eq:Majorana_Ham}
\eea
Eq. (\ref{eq:Majorana_Ham}) can alternatively be written in terms of a spinless complex fermion $c_j,c_j^\dagger$ as
\bea
H_{e.s.}&=&\frac{1}{4}\Gamma_1\sum_n [-i(c_{2n}^\dagger c_{2n-1}^\dagger-c_{2n-1}c_{2n})\nn\\
&& +i(c_{2n-1}^\dagger c_{2n}-c_{2n}^\dagger c_{2n-1}) \nn\\
&&-(c_{2n+1}^\dagger c_{2n}^\dagger+c_{2n}c_{2n+1})\nn\\
&&+(c_{2n+1}^\dagger c_{2n}+c_{2n}^\dagger c_{2n+1})].
\label{eq:Hes_realspace}
\eea
Introducing the following half-lattice Fourier transformations,
\bea
c_{A,k}^\dagger&=&\frac{1}{\sqrt{L/2}}\sum_n c_{2n}^\dagger e^{ik\cdot 2n}\nn\\
c_{B,k}^\dagger&=&\frac{1}{\sqrt{L/2}}\sum_n c_{2n+1}^\dagger e^{ik\cdot (2n+1)},
\label{eq:Fourier_c}
\eea
Eq. (\ref{eq:Hes_realspace}) can be cast into the following Bogoliubov-de Gennes (BdG) form 
\bea
H_{e.s.}&=& \sum_k \Psi^\dagger_k H_k \Psi_k,
\eea
in which $\Psi_k^\dagger$ is a four-component row vector defined as
\bea
\Psi_k^\dagger = (c_{A,k}^\dagger,c_{B,k}^\dagger, c_{A,-k},c_{B,-k}),
\eea
and $H_k$ is a $4\times 4$ matrix defined as 
\begin{flalign}
&H_k=\frac{\Gamma_1}{4}\times\nn\\
&\left(\begin{array}{cccc}
0 & 1-ie^{2ik} & 0 & 1-ie^{2ik}\\
1+ie^{-2ik} & 0 & -1+ie^{-2ik} & 0  \\
0 & -1-ie^{2ik} & 0 & -1-ie^{2ik}\\
1+ie^{-2ik} & 0 & -1+ie^{-2ik} & 0
\end{array}\right).
\end{flalign}
Notice that for a periodic chain with $L$ sites,
the summation over $k$ runs from $0$ to $\pi$ with spacing $2\pi/L$,
hence the number of distinct wavevectors is $L/2$,
which is the origin of the $\sqrt{2/L}$ prefactors in Eq. (\ref{eq:Fourier_c}).

The four eigenvalues of $H_k$ can be obtained as
\begin{flalign}
&E_{k,1}=E_{k,2}=0,\nn \\
&E_{k,3}=E_{k,4}=\sqrt{2}\Gamma,
\label{eq:es_spectrum}
\end{flalign}
in which $\Gamma_1=2\Gamma$ is used. 
In particular, the two zero modes with wavevector $k$ corresponding to $E_{k,1}$ and $E_{k,2}$ are given by 
\begin{flalign}
&\alpha_{k}^\dagger= \frac{1}{\sqrt{2}}[e^{-i\pi/4}\cos(\frac{\pi}{4}+k) c_{A,k}^\dagger+e^{i\pi/4}\cos(\frac{\pi}{4}-k) c_{A,-k}],\nn\\
&\beta_{k}^\dagger= \frac{i}{\sqrt{2}}(- c_{B,k}^\dagger +c_{B,-k}).
\end{flalign}
It can easily seen that 
\bea
\alpha_{k}^\dagger&=&\alpha_{-k},\nn\\
\beta_{k}^\dagger&=&\beta_{-k},
\eea
hence the following two sets of operators are hermitian operators with zero energies,
\bea
\alpha_{2n}&=&\frac{1}{\sqrt{L/2}}\sum_k \alpha_{k}^\dagger e^{ik\cdot 2n},\nn\\
\beta_{2n}&=&\frac{1}{\sqrt{L/2}}\sum_k \beta_{k}^\dagger e^{ik\cdot 2n},
\eea
namely, they are Majorana zero modes of the system. 

\begin{figure}[ht]
\begin{center}
\includegraphics[width=8cm]{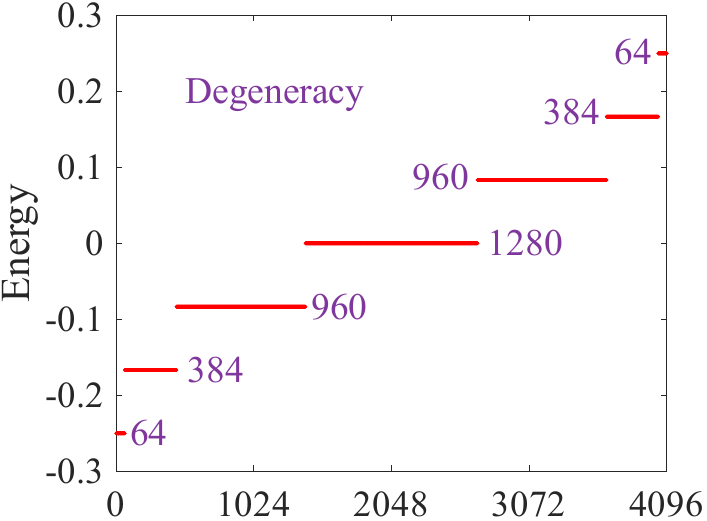}
\caption{Energy per site of a periodic chain with parameters $(K=0,\Gamma=D_M=1/\sqrt{2})$ obtained from exact diagonalization on $L=12$ sites. 
The numbers in magenta denote the degeneracies of the corresponding equally spaced energy levels. 
The predicted energy spacing per site is $\sqrt{2}\Gamma/L=1/L$, which is consistent with the numerical results. 
} \label{fig:degeneracy}
\end{center}
\end{figure}

Since the system has two sets of Majorana operators, each containing $L/2$ elements,
there are in total $L$ Majorana zero modes.
As a result, the system has a $2^{L/2}$-fold ground state degeneracy.
Also notice that all quasi-particle excitations at different wavevectors form  flat bands,  having the same energy $\sqrt{2}\Gamma$ as can be seen from $E_{k,3}$  and  $E_{k,4}$ from Eq. (\ref{eq:es_spectrum}),
meaning the energy spectrum of the system is equally spaced. 
Both the exponential ground state degeneracy and the equal spacing of energy levels have been verified by the exact diagonalization method on a system of $L=12$ sites as shown in Fig. \ref{fig:degeneracy}.
Because of the exponentially large ground state degeneracy, the usual perturbative analytical treatment fails in the vicinity of the $D_M=\pm \Gamma$ points in the phase diagram.
In addition, a huge number of low energy states makes numerical calculations very difficult and costly. 
Hence, it is not unexpected that both analytical and numerical difficulties will be encountered in the parameter region close to the $D_M=\pm \Gamma$ points as will be discussed in Sec. \ref{subsubsec:non_SU2_pt4}. 

\section{Emergent SU(2)$_1$ conformal symmetry in the spin-1/2 Kitaev-Gamma-DM chain}
\label{sec:KGDM_SU2_1}

In this section, we analytically study the spin-1/2 Kitaev-Gamma chain with a bond-dependent DM interaction, which can be induced by an electric field along $z$-direction.
We first briefly  summarize the main results of this section. 
Fig. \ref{fig:phase_KGDM} shows the plot of half of the unit sphere of the parameter space 
in the region $\theta\in(0,\pi/2)$ and $\phi\in(0,2\pi)$, where $\theta$ and $\phi$ as defined in Eq. (\ref{eq:parametrize_KGDM}),
and the other half of the sphere is not shown because of the equivalence in Eq. (\ref{eq:equiv0}). 
The phase diagram in Fig.  \ref{fig:phase_KGDM}  is reflection symmetric with respect to the horizontal line connecting $K$ and $-K$ points,
which is a consequence of the equivalence $(K,\Gamma,D_M)\simeq (K,-\Gamma,D_M)$ as a result of Eqs. (\ref{eq:equiv0},\ref{eq:equiv1}), 
We note that in Fig. \ref{fig:phase_KGDM}, the value of $\theta$ corresponds to the radial length, and $\phi$ is represented in a circular manner.

We will analytically demonstrate that there is an emergent SU(2)$_1$ conformal symmetry at low energies in the ``Emergent SU(2)$_1$ I" region in Fig. \ref{fig:phase_KGDM}, which is shown by perturbing the Kitaev-Gamma model and treating the DM interaction as a perturbation. 
Then it can be established  that the regions marked with ``Emergent SU(2)$_1$ II" and ``Emergent SU(2)$_1$ III" also have emergent SU(2)$_1$ conformal symmetries at low energies,
by applying the equivalences $(K,\Gamma,D_M)\simeq (K,-\Gamma,D_M)$ (from Eqs. (\ref{eq:equiv0},\ref{eq:equiv1})) and  
$(K,\Gamma,D_M)\simeq (-K,D_M,\Gamma)$ (from Eqs. (\ref{eq:equiv1},\ref{eq:equivalence})) to the ``Emergent SU(2)$_1$ I" phase, respectively. 
The phase boundaries in Fig. \ref{fig:phase_KGDM} are determined by our large-scale DMRG numerical simulations as discussed in Sec. \ref{subsec:phase_transition}.

\subsection{Low energy field theory}
\label{subsec:Low_field_theory}

In this subsection, by treating the DM term as a perturbation, we derive the low energy field theory of the model in the ``Emergent SU(2)$_1$ I" phase in Fig. \ref{fig:phase_KGDM}.
We start with a brief review of the emergent SU(2)$_1$ CFT in the Kitaev-Gamma model without DM interaction. 


When $K=\Gamma_1=\Gamma_2<0$, the system in Eq. (\ref{eq:Ham_rot}) reduces to the AFM Heisenberg model, which represents a hidden SU(2) symmetric point.  
It is known that at low energies,  the spin-1/2 AFM Heisenberg model is described by the SU(2)$_1$ Wess-Zumino-Witten (WZW) model.
The Hamiltonian density is of the Sugawara form 
\bea
\mathcal{H}=\frac{2\pi}{3}v (\vec{J_L}\cdot \vec{J_L}+\vec{J_R}\cdot \vec{J_R})
\eea
with an additional marginally irrelevant term $-u\vec{J_L}\cdot \vec{J_R}$, 
in which $v$ is the spin velocity; $u>0$ is the coupling constant of the marginally irelevant term $\vec{J_L}\cdot \vec{J_R}$;
and $\vec{J_L}$ and $\vec{J_R}$ defined by 
\bea
\vec{J_L}&=&-\frac{1}{4\pi}\text{tr} [(\partial_z g) g^\dagger \vec{\sigma}] \nn\\
\vec{J_R}&=&\frac{1}{4\pi}\text{tr} [g^\dagger (\partial_{\bar{z}} g) \vec{\sigma}],
\label{eq:J_LR}
\eea
are the left and right WZW currents, respectively, 
where the SU(2) matrix $g$ is the WZW primary field, $\sigma^{\alpha}$ ($\alpha=x,y,z$) are the three Pauli matrices,
and $z=\tau+ix$ ($\bar{z}=\tau-ix$) is the holomorphic (anti-holomorphic) coordinate in the imaginary time formalism.
This hidden SU(2) symmetric AFM point provides a starting point for a field theory perturbation in the regions nearby.
For later convenience, we define 
\bea
\epsilon&=&\text{tr}(g), \nn\\
N^\alpha&=&i \text{tr}(g\sigma^\alpha).
\label{eq:epsilon_N1}
\eea

When $D_M=0$ but $K\neq \Gamma$, the system in Eq. (\ref{eq:Ham_rot}) reduces to the Kitaev-Gamma model,
and it has been shown in Ref. \onlinecite{Yang2020} that the system is described by the SU(2)$_1$ WZW model at low energies.
The spin operators are related to the SU(2)$_1$ low energy fields by the nonsymmorphic nonabelian bosonization formula:
\bea
\frac{1}{a}S_{i+3n}^\alpha=D^\alpha_{i} (J_L^\alpha+ J_R^\alpha)+C^\alpha_{i} \frac{1}{\sqrt{a}} (-)^{n} N^\alpha,
\label{eq:nonsymmorphic_bosonize}
\eea
in which $i=1,2,3$; $\alpha=x,y,z$;
and 
\begin{flalign}
&D_1^z=D_2^y=D_3^x (=D_1),\nn\\
&D_1^x=D_2^z=D_3^y=D_1^y=D_2^x=D_3^z (=D_2),\nn\\
&C_1^z=C_2^y=C_3^x (=C_1),\nn\\
&C_1^x=C_2^z=C_3^y=C_1^y=C_2^x=C_3^z (=C_2).
\end{flalign}


Next we consider the case with a nonzero DM interaction. 
When $D_M\neq 0$, in the six-sublattice rotated frame, 
the perturbation is 
\bea
\Delta H^\prime=D_M\sum_{<ij>\in\gamma\,\text{bond}}(-S_i^\alpha S_j^\alpha+ S_i^\beta S_j^\beta),
\label{eq:DeltaH_D}
\eea
in which $\gamma\in\{x,\bar{z},y,\bar{x},z,\bar{y}\}$;
 $(\alpha\beta\gamma)$ form a right-handed coordinate system for $\gamma\in\{x,y,z\}$, and  form a left-handed system when $\gamma\in\{\bar{x},\bar{y},\bar{z}\}$.
 
By treating $D_M$ as a perturbation on the Kitaev-Gamma model,  we can use Eq. (\ref{eq:nonsymmorphic_bosonize}) to express $\Delta H^\prime$ in  Eq. (\ref{eq:DeltaH_D}) in terms of the SU(2)$_1$ low energy degrees of freedom.
The detailed calculations are included in Appendix \ref{app:derivation}.
Here we only quote the result
\begin{flalign}
\Delta H^\prime = -D_M(C_1D_2-C_2D_1)\int dx :(\vec{J}_L+\vec{J}_R)\cdot\vec{N}:.
\label{eq:Delta_H_JN}
\end{flalign}
Clearly, the $:(\vec{J}_L+\vec{J}_R)\cdot\vec{N}:$ term does not vanish when $C_1D_2-C_2D_1\neq 0$,
which is the case except $K=\Gamma$.
In fact, when $K=\Gamma$, the $:(\vec{J}_L+\vec{J}_R)\cdot\vec{N}:$ term only appears when the system contains second nearest neighbor interactions, as discussed in details in Appendix \ref{app:2nd_quadrupole}.
To summarize, the full low energy Hamiltonian is given by
\bea
\mathcal{H}&=&\frac{2\pi v}{3} \int dx :(\vec{J}_L\cdot \vec{J}_L+\vec{J}_R\cdot \vec{J}_R):-u\int dx \vec{J}_L\cdot \vec{J}_R\nn\\
&&-D_M(C_1D_2-C_2D_1) \int dx :(\vec{J}_L+\vec{J}_R)\cdot \vec{N}:.
\label{eq:low_energy_theory2}
\eea

\subsection{Emergent SU(2)$_1$ conformal symmetry}
\label{sec:emergent_SU2}

At first sight, the system opens a gap at low energies since the scaling dimension of  $:(J^\alpha_L+J^\alpha_R) N^\alpha:$ is $3/2$ which is a relevant operator.
However,  $:(J^\alpha_L+J^\alpha_R) N^\alpha:$ is a total derivative in the SU(2)$_1$ WZW model,  hence it has no effect in the low energy Hamiltonian.
There is a quick way to see this. 
In abelian bosonization, we have (up to  overall constant factors) \cite{Giamarchi2004}
\bea
J^z&=&-\frac{1}{\pi}\nabla \varphi,\nn\\
N^z&=& \frac{1}{\pi } \cos(2\varphi),
\eea
in which $\varphi$ is the field in the Luttinger liquid Hamiltonian
\bea
H_{LL}=\frac{v}{2\pi} \int dx [\kappa^{-1} (\nabla \varphi)^2 +\kappa (\nabla \theta)^2].
\eea
It is clear that $J^zN^z=-\frac{1}{2\pi^2}\nabla \sin(2\varphi)$ is a total derivative.
Since the SU(2)$_1$ WZW model has SU(2) symmetry, $:J^xN^x:$ and $:J^yN^y:$ must also be total derivatives.
In fact, based on the OPE relations in the SU(2)$_1$ WZW model, a rigorous proof of $:(J^\alpha_L+J^\alpha_R) N^\alpha:$ being a total derivative can be given as  discussed in Appendix \ref{app:total_derivative}, which shows that
\bea
:(\vec{J}_L+\vec{J}_R) \cdot \vec{N}:=-3\nabla \epsilon.
\eea

Dropping the $:(J^\alpha_L+J^\alpha_R) N^\alpha:$ term, Eq. (\ref{eq:low_energy_theory2}) becomes
\begin{flalign}
\mathcal{H}&=&\frac{2\pi v}{3} \int dx :(\vec{J}_L\cdot \vec{J}_L+\vec{J}_R\cdot \vec{J}_R):-u\int dx \vec{J}_L\cdot \vec{J}_R,
\label{eq:low_energy_theory_2}
\end{flalign}
which remains to have an emergent SU(2)$_1$ conformal symmetry even for nonzero $D_M$,
as along as $D_M$ is small enough (i.e., $D_M\ll \sqrt{K^2+\Gamma^2}$) such that a perturbative analysis is valid.

\section{Symmetry analysis }
\label{sec:low_theory_DM_z}

The analysis in Sec. \ref{subsec:Low_field_theory} is only based on a first order perturbative treatment.
To fully confirm that except $:(J^\alpha_L+J^\alpha_R) N^\alpha:$, there is no other additional relevant or marginal term (i.e., having scaling dimension less than or equal to two) in the low energy field theory compared with the $D_M=0$ case, we perform a symmetry analysis to analyze all the symmetry allowed terms among the relevant and marginal operators. 
This is a proof for the emergent SU(2)$_1$ conformal symmetry for small enough $D_M$. 
For simplification of notations,  we will write the normal ordered product $:AB:$ as $AB$ in this section.

\subsection{Nonsymmorphic symmetry group}

We first analyze the symmetry group of the model in the six-sublattice rotated frame. 
The Hamiltonian $H^\prime_z$ in Eq. (\ref{eq:Ham_rot}) is invariant under the following symmetry operations,
\begin{eqnarray}
1.&T &:  (S_i^x,S_i^y,S_i^z)\rightarrow (-S_{i}^x,-S_{i}^y,-S_{i}^z)\nn\\
2.& R_a^{-1}T_{2a}&:  (S_i^x,S_i^y,S_i^z)\rightarrow (S_{i+2}^y,S_{i+2}^z,S_{i+2}^x)\nn\\
3.&R_I I&: (S_i^x,S_i^y,S_i^z)\rightarrow (-S_{4-i}^z,-S_{4-i}^y,-S_{4-i}^x)\nn\\
4.&R(\hat{x},\pi) &:  (S_i^x,S_i^y,S_i^z)\rightarrow (S_i^x,-S_i^y,-S_i^z)\nn\\
5.&R(\hat{y},\pi) &:  (S_i^x,S_i^y,S_i^z)\rightarrow (-S_i^x,S_i^y,-S_i^z)\nn\\
6.&R(\hat{z},\pi) &:  (S_i^x,S_i^y,S_i^z)\rightarrow (-S_i^x,-S_i^y,S_i^z),
\label{eq:symmetries_Ez}
\end{eqnarray}
in which $R(\hat{n},\phi)=e^{i\phi\sum_j \vec{S}_j\cdot \hat{n}}$ denotes a global spin rotation around $\hat{n}$-direction by an angle $\phi$;
$T_{na}$ represents the translation operator by $n$ sites;
$R_a$ is the rotation around $(1,1,1)$-direction by $2\pi/3$;
and $R_I$ is a $\pi$-rotation around the $(1,0,-1)$-direction.
The symmetry group $G$ is generated by the operations in Eq. (\ref{eq:symmetries_Ez}), i.e., 
\bea
G_z=\mathopen{<}T, R_a^{-1}T_{2a},R_I I,R(\hat{x},\pi),R(\hat{y},\pi),R(\hat{z},\pi)\mathclose{>}.
\label{eq:Gz}
\eea

We note that $G_z$ is very similar to the symmetry group $G_d$ of the dimerized Kitaev-Gamma chain discussed in Ref. \onlinecite{Yang2022}.
The symmetry operations of $G_z$ and $G_d$ overlap except the inversion symmetry: The inversion center for the inversion operation in $G_z$ is a site, whereas it is located at bond center for $G_d$.
However,  this difference has notable physical effects:
The dimerized spin-1/2 Kitaev-Gamma model is in a disordered phase with a nonzero spin gap,  whereas the spin-1/2 Kitaev-Gamma-DM model is gapless having an emergent SU(2)$_1$ conformal symmetry at low energies. 
It is worth to mention that the group structure of $G_z$ satisfies $G_z/\mathopen{<}T_{6a}\mathclose{>}\simeq O_h$ where $O_h$ is the full octahedral group, which can be proved in a similar way as what is done for $G_d$ in Ref. \onlinecite{Yang2022}.

\subsection{Symmetry analysis of the low energy Hamiltonian}

Sec. \ref{subsec:Low_field_theory} demonstrates the existence of the $(J^\alpha_L+J^\alpha_R) N^\alpha$ term. 
In this section, by exploiting a symmetry analysis, we prove that there is no other relevant operators in the low energy field theory.
 

The symmetry transformation properties of the WZW fields under spin rotations, time reversal, and inversion operations are summarized in Appendix \ref{app:transform_WZW}.
A complete  analysis of all relevant and marginal operators in the SU(2)$_1$ WZW model based on symmetry considerations is as follows.

1) $\epsilon$ changes sign under $R_II$, hence forbidden. 

2) $N^\alpha$ ($\alpha=x,y,z$) changes sign under $T$, hence forbidden.

3) $J^\alpha$ ($\alpha=x,y,z$) changes sign under $R(\hat{\beta},\pi)$ where $\beta\neq \alpha$, hence forbidden. 

4) $J^\alpha_L\epsilon,J^\alpha_R\epsilon$ change sign under $R(\hat{\beta},\pi)$ where $\beta\neq \alpha$, hence forbidden. 

5) Within $J^\alpha_LN^\beta,J^\alpha_RN^\beta$, the only allowed interaction is 
$(\vec{J}_L+\vec{J}_R)\cdot \vec{N}$ because of the $O_h$ symmetry. 

6) For $J_L^\alpha J_R^\beta$, the only allowed term is
$\vec{J}_L \cdot\vec{J}_R$ because of the $O_h$ symmetry. 

According to the above analysis, the symmetry allowed low energy Hamiltonian is given by
\bea
\mathcal{H}&=&\frac{2\pi v}{3} \int dx (\vec{J}_L\cdot \vec{J}_L+\vec{J}_R\cdot \vec{J}_R)-u\int dx \vec{J}_L\cdot \vec{J}_R\nn\\
&&+\lambda \int dx (\vec{J}_L+\vec{J}_R)\cdot \vec{N}.
\label{eq:low_energy_theory}
\eea
Indeed, $(J^\alpha_L+J^\alpha_R) N^\alpha$ is the only symmetry allowed relevant operator in the low energy Hamiltonian.
The emergent SU(2)$_1$ conformal symmetry then follows from the observation that the $(J^\alpha_L+J^\alpha_R) N^\alpha$ term is a total derivative 
as discussed in Sec. \ref{sec:emergent_SU2}. 
It is noteworthy to mention that the low energy Hamiltonian in Eq. (\ref{eq:low_energy_theory}) is SU(2) invariant although the symmetry group is discrete and nonsymmorphic.

\subsection{Nonsymmorphic nonabelian bosonization formulas}
\label{sec:nonsymmor_bosonize}

In this subsection, we derive the nonsymmorphic nonabelian bosonization formulas for the Kitaev-Gamma-DM model based on a symmetry analysis. 
These formulas only respect the exact nonsymmorphic symmetry group $G_z$ in Eq. (\ref{eq:Gz}), and break the emergent SU(2) symmetry. 

In general, the local spin operators are related to the SU(2)$_1$ low energy degrees of freedom via the following relations,
\bea
S_j^\alpha&=& \sum_\beta[ D_{L,j}^{\alpha\beta} J^\beta_L+D_{R,j}^{\alpha\beta} J^\beta_R +(-)^j C_j^{\alpha\beta} N^\beta],
\label{eq:bosonizaion_1}
\eea
in which $\alpha,\beta=x,y,z$.
We analyze how symmetry constrains the coefficients in Eq. (\ref{eq:bosonizaion_1}).
First, since time reversal symmetry switches the left and right movers, we have 
\bea
D_{L,j}^{\alpha\beta}=D_{R,j}^{\alpha\beta}=D_{j}^{\alpha\beta}.
\eea
Second, since $R(\hat{\alpha},\pi)$ ($\alpha=x,y,z$) leaves the system invariant, it is clear that the cross coefficients $D_j^{\alpha\beta}$ and $C_j^{\alpha\beta}$ for $\alpha\neq \beta$ vanish in Eq. (\ref{eq:bosonizaion_1}).
Third, the symmetry operation $R_a^{-1}T_{2a}$ requires
\begin{flalign}
&D_1^{xx}=D_3^{yy}=D_5^{zz}=D_2\nn\\
&D_1^{yy}=D_3^{zz}=D_5^{xx}=D_3\nn\\
&D_1^{zz}=D_3^{xx}=D_5^{yy}=D_1\nn\\
&D_2^{xx}=D_4^{yy}=D_6^{zz}=D_2^\prime \nn\\
&D_2^{yy}=D_4^{zz}=D_6^{xx}=D_1^\prime\nn\\
&D_2^{zz}=D_4^{xx}=D_6^{yy}=D_3^\prime.
\end{flalign}
Fourth, the $R_II$ symmetry requires
\bea
D_2=D_3,~D_2^\prime=D_3^\prime.
\eea
Similar relations hold for the $C_j^{\alpha\beta}$'s coefficients. 
The explicit expressions of the nonsymmorphic nonabelian bosonization formulas are included in Appendix \ref{app:bosonization}.

\begin{figure*}[ht]
\begin{center}
\includegraphics[width=17cm]{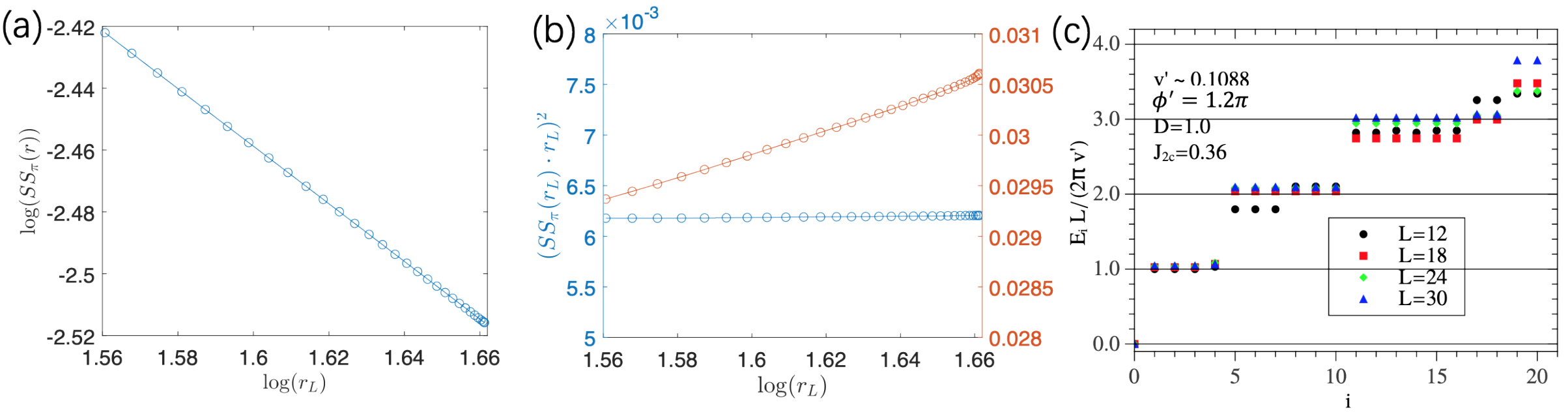}
\caption{(a) Averaged correlation function $\frac{1}{3}\sum_{\alpha=x,y,z} S^{\alpha\alpha}_\pi(r)$ as a function of $r_L=\frac{L}{\pi}\sin(\frac{\pi r}{L})$ on a log-log scale for $J_2=0$,
(b) $[\frac{1}{3}\sum_\alpha S^{\alpha\alpha}_\pi(r)\cdot r_L]^2$ as a function of $\log r_L$ for $J_2=0$ (orange line) and $J_2=0.36$ (blue line),
(c) excitation energies of low energy multiplets at $J_2=0.36$ for various system sizes.
In (a,b), $\frac{1}{3}\sum_{\alpha=x,y,z} S^{\alpha\alpha}_\pi(r)$ in the vertical axes are denoted as $SS_\pi(r)$ for short. 
In (a,b,c), the parameters in Eq. (\ref{eq:new_parametrization}) are taken as $\phi^\prime=1.2\pi$, $D_M=1$. 
In (a,b), DMRG numerics are performed on a system of $L=144$ sites with periodic boundary conditions where the bond dimension $m$ and truncation error $\epsilon$ are take as $m=1200$ and $\epsilon=10^{-9}$.
In (c), exact diagonalization is performed on systems sizes $L=12,18,24,30$ with periodic boundary conditions.  
} 
\label{fig:correlation}
\end{center}
\end{figure*}

With these bosonization formulas, one can calculate any low energy property of the system using the SU(2)$_1$ WZW model.
For example, the static correlation function $S^{\alpha\alpha}(r)=\langle S_1^\alpha S_{1+r}^\alpha \rangle$ can be derived in the $r\gg 1$ limit as
\begin{flalign}
&\langle S^{\alpha}_i S^{\beta}_{j+6n}  \rangle =\nn\\
& \delta_{\alpha \beta} \big[ -D_{i}^{\alpha\alpha} D_{j}^{\alpha\alpha}\frac{1}{r^2} +(-)^r C_{i}^{\alpha\alpha} C_{j}^{\alpha\alpha}\frac{\ln^{1/2}(r/r_0)}{r} \big].
\label{eq:modified_correlation}
\end{flalign}
To derive Eq. (\ref{eq:modified_correlation}),
by properly normalizing the WZW current operators and the primary field, 
the following formulas are used 
\bea
\langle J^\alpha(0)J^\beta(r)\rangle &=&\delta_{\alpha\beta}\frac{1}{r^2},\nn\\
\langle N^\alpha(0)N^\beta(r)\rangle &=&\delta_{\alpha\beta}\frac{[\ln (r/r_0)]^{1/2}}{r},
\label{eq:correlation_WZW}
\eea
in which $J^\alpha=J^\alpha_L+J^\alpha_R$; 
$\langle ... \rangle$ represents the expectation value over the ground state;
the time variable is taken as zero; 
the arguments in $J^\alpha$ and $N^\alpha$ are spatial coordinates; 
the logarithmic correction in $\langle N^\alpha(0)N^\beta(r)\rangle$ comes from the marginally irrelevant operator $-u\int \vec{J}_L\cdot \vec{J}_R$ in the low energy theory;
and $r_0$ is an ultraviolet cutoff, 
which is of the same order  as the lattice constant. 
Notice that because of the six-site periodicity in Eq. (\ref{eq:modified_correlation}), in addition to the gapless wavevectors $0$ and $\pi$,
the system is also gapless at wavevectors $\pm\pi/3$, $\pm2\pi/3$,
which is a consequence of the nonsymmorphic bosonization formulas in Eq. (\ref{eq:bosonizaion_1}).
Finally, we note that the origin of the nonsymmorphic bosonization coefficients $C_i$, $D_i$, $C_i^\prime$, $D_i^\prime$ ($i=1,2$) can be understood from wavefunction renormalization effects at the ``Planck scale" of the lattice as discussed in details in Ref. \onlinecite{Yang2020c}.

\section{Numerical results}
\label{sec:numerics}

In this section, we present numerical evidences for the emergent SU(2)$_1$ conformal symmetry in the 1D spin-1/2 Kitaev-Gamma-DM  model and determine the ranges of the emergent SU(2)$_1$  phases.
DMRG numerical simulations are performed in the six-sublattice rotated frame defined in Eq. (\ref{eq:6rotation}).

\subsection{Numerical evidence for emergent SU(2)$_1$ conformal symmetry}
\label{subsec:SU2_1}

We provide multiple numerical evidence for the emergent SU(2)$_1$ conformal symmetry.
Throughout this subsection, the following parametrization will be used 
\bea
K&=&\cos(\phi^\prime),\nn\\
\Gamma&=&\sin(\phi^\prime),\nn\\
D_M&=&D_M.
\label{eq:new_parametrization}
\eea
A representative value of $\phi^\prime=1.2\pi$ is taken in DMRG numerics in this subsection. 
Comparing Eq. (\ref{eq:new_parametrization}) with Eq. (\ref{eq:parametrize_KGDM}),
it can be easily seen that the relation between the two parametrizations  is $\phi^\prime=\phi$, $\theta=\arcsin(1/\sqrt{1+D_M^2})$.
The line $\phi^\prime=1.2\pi$ (i.e., $\phi=1.2\pi$) has been marked in Fig. \ref{fig:phase_KGDM} as the purple dashed line for reader's convenience. 


\subsubsection{Correlation functions}
\label{subsec:correlation}

As discussed in Ref. \onlinecite{Yang2020}, $\phi^\prime=1.2\pi$ lies in the emergent SU(2)$_1$ phase of the Kitaev-Gamma model. 
According to previous analysis, we expect that by adding a small $D_M$ at $\phi^\prime=1.2\pi$, the system remains to have an emergent SU(2)$_1$ conformal symmetry at low energies. 
It turns out  that even a value of $D_M$ as large as $D_M=1.0$ works,
which is denoted as the solid green circle in Fig. \ref{fig:phase_KGDM}.
Fig. \ref{fig:correlation} (a) shows $\frac{1}{3}\sum_{\alpha=x,y,z} S^{\alpha\alpha}_\pi(r)$ as a function of $r_L$ on a log-log scale,
in which  $\phi^\prime=1.2\pi$, $D_M=1$ as defined in Eq. (\ref{eq:new_parametrization});
$S^{\alpha\alpha}_\pi(r)$ is the $\pi$-wavevector oscillating  component of the correlation function $\langle S^{\alpha}_1 S^{\alpha}_{r} \rangle$ as a function of $r$;
$r_L=\frac{L}{\pi}\sin(\frac{\pi r}{L})$ in accordance with CFT in finite size systems \cite{DiFrancesco1997}; 
and DMRG numerics are performed on a system of $L=144$ sites with periodic boundary conditions where the bond dimension $m$ and truncation error $\epsilon$ are taken as $m=1200$ and $\epsilon=10^{-9}$.
The slope of the line in Fig. \ref{fig:correlation} (a) is extracted to be $-0.935$ by assuming a linear relation,
which is very close to $-1$ as predicted by the SU(2)$_1$ CFT. 
In fact, the deviation from $-1$ originates from the logarithmic correction in Eq. (\ref{eq:modified_correlation}). 

To further study the logarithmic correction, we plot $[\frac{1}{3}\sum_\alpha S^{\alpha\alpha}_\pi(r)\cdot r_L]^2$ as a function of $\log r_L$ as shown by the orange line in Fig. \ref{fig:correlation} (b).
It can be observed that the line is linear with a non-vanishing slope, indicating a logarithmic correction factor with a $1/2$  power in $S^{\alpha\alpha}_\pi(r)$. 
In fact, the logarithmic correction can be killed by introducing a second nearest neighbor Heisenberg term \cite{Affleck1988} in the six-sublattice rotated frame.
By introducing a $J_2$ term into Eq. (\ref{eq:Ham_rot}), we consider the following Hamiltonian, 
\bea
H^\prime_2&=&\sum_{\langle ij \rangle\in\gamma\,\text{bond}}\big(-KS_i^\gamma S_j^\gamma - \Gamma_1 S_i^\alpha S_j^\alpha-\Gamma_2 S_i^\beta S_j^\beta\big),\nn\\
&&+J_2\sum_{i} \vec{S}_i\cdot \vec{S}_{i+2}.
\label{eq:Ham_rot2}
\eea
The $J_2$ term renormalizes the coupling constant $u$ in the $\vec{J}_L\cdot \vec{J}_R$ term in Eq. (\ref{eq:low_energy_theory_2}). 
When $J_2$ is  at a  critical value $J_{c2}$, the coupling $u$ vanishes and the logarithmic correction is killed. 
The blue line in Fig. \ref{fig:correlation} (b) shows $[\frac{1}{3}\sum_\alpha S^{\alpha\alpha}_\pi(r)\cdot r_L]^2$ as a function of $\log r_L$ when $J_2$ is chosen as $J_2=0.36$.
Clearly, the blue line has a significantly reduced slope compared with the orange line in the same figure,
hinting a critical value  $J_{c2}$ very close to $0.36$.
Furthermore, we have calculated the excitation energies of several low energy multiplets at $J_2=0.36$ for several system sizes using exact diagonalization as shown in Fig. \ref{fig:correlation} (c).
As can be seen in Fig. \ref{fig:correlation} (c), as the system size is increased,
the energy plateaus are approaching the conformal towers of the SU(2)$_1$ WZW model, 
indicating an absence of the $\vec{J}_L\cdot \vec{J}_R$ term in the low energy theory (whose presence in general spoils the structure of the conformal tower in a logarithmic manner).

\subsubsection{Central charge}
\label{subsec:central_charge}

\begin{figure}[ht]
\begin{center}
\includegraphics[width=7cm]{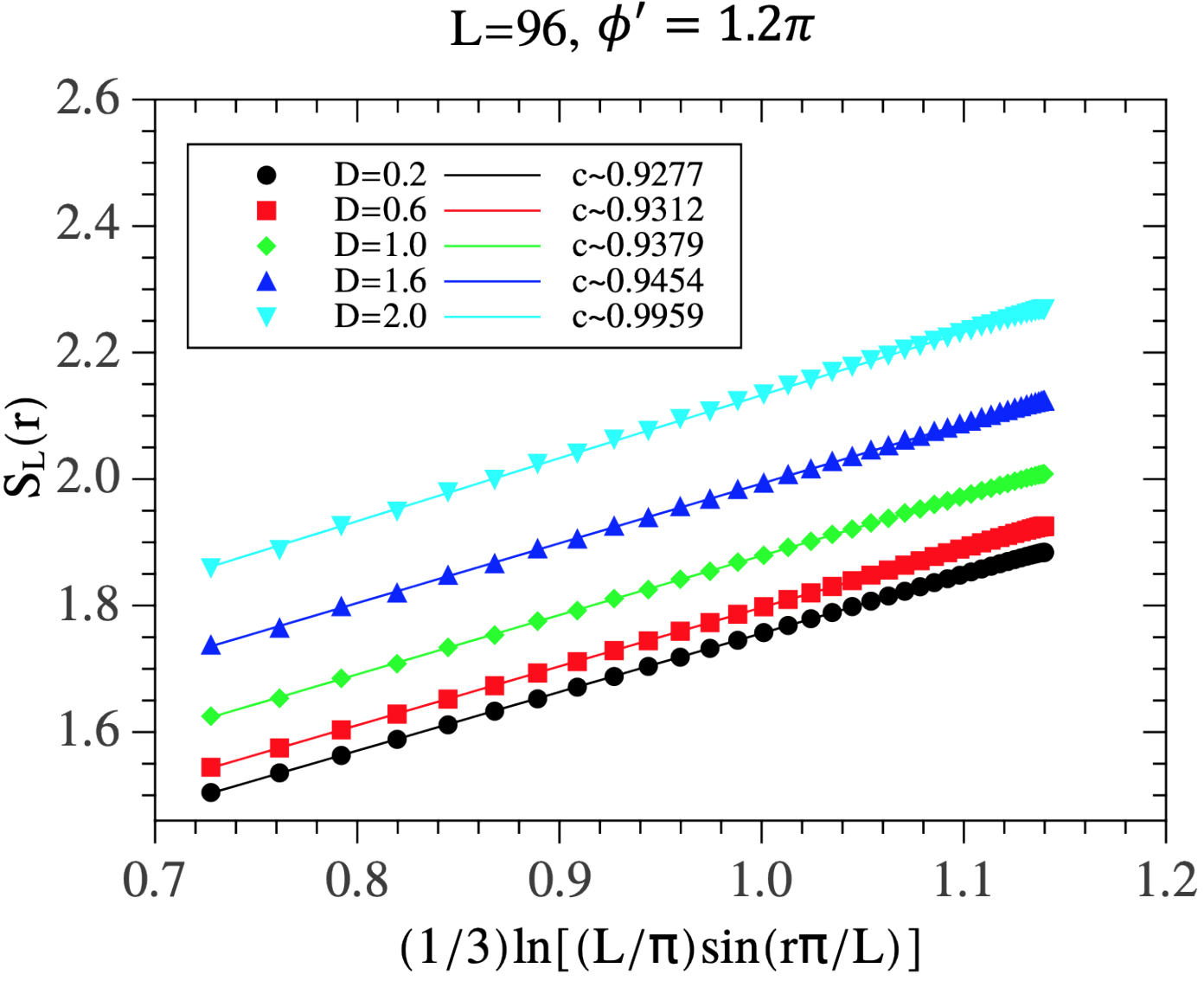}
\caption{
$S_L(r)$ as a function of $\frac{1}{3} \ln \left[ \frac{L}{\pi} \sin\left( \frac{\pi r}{L} \right) \right]$ for a variety of $D_M$ at $\phi^\prime=1.2\pi$, where the parametrization is defined in Eq. (\ref{eq:new_parametrization}).
DMRG numerics are performed on a system of $L=96$ sites with a periodic boundary condition where the bond dimension $m$ and truncation error $\epsilon$ are taken as $m=1200$ and $\epsilon=10^{-8}$.
} 
\label{fig:DM_central_charge2}
\end{center}
\end{figure}

We further study the central charge (denoted as $c$) of the system in the ``Emergent SU(2)$_1$ I" phase in Fig. \ref{fig:phase_KGDM},
which is predicted to be $c=1$ in the SU(2)$_1$ CFT. 
DMRG numerics are performed on a system of $L=96$ sites with periodic boundary conditions where the bond dimension $m$ and truncation error $\epsilon$ are taken as $m=1200$ and $\epsilon=10^{-8}$.
The parameter $\phi^\prime$ in Eq. (\ref{eq:new_parametrization}) is fixed to be $\phi^\prime=1.2\pi$, and $D_M$ is tuned. 
We compute the entanglement entropy $S_L(x)$ of a subregion $x$ to extract the value of the central charge. 
Conformal field theory predicts the following scaling of the entanglement entropy \cite{Calabrese2009}
\begin{align}
    S_L(r) = \frac{c}{3} \ln \left[ \frac{L}{\pi} \sin\left( \frac{\pi r}{L} \right) \right] + \cdots,
    \label{eq:entanglement}
\end{align}
in which ``$\cdots$" denotes subleading terms. 
Fig. \ref{fig:DM_central_charge2} shows $S_L(r)$ vs. $\frac{1}{3} \ln \left[ \frac{L}{\pi} \sin\left( \frac{\pi r}{L} \right) \right]$ for a variety of $D_M$ at $\theta=1.2\pi$.
It is clear from Fig. \ref{fig:DM_central_charge2} that the slopes are all very close to $1$, indicating $c=1$ in accordance with Eq. (\ref{eq:entanglement}). 

\begin{figure}[ht]
\begin{center}
\includegraphics[width=7cm]{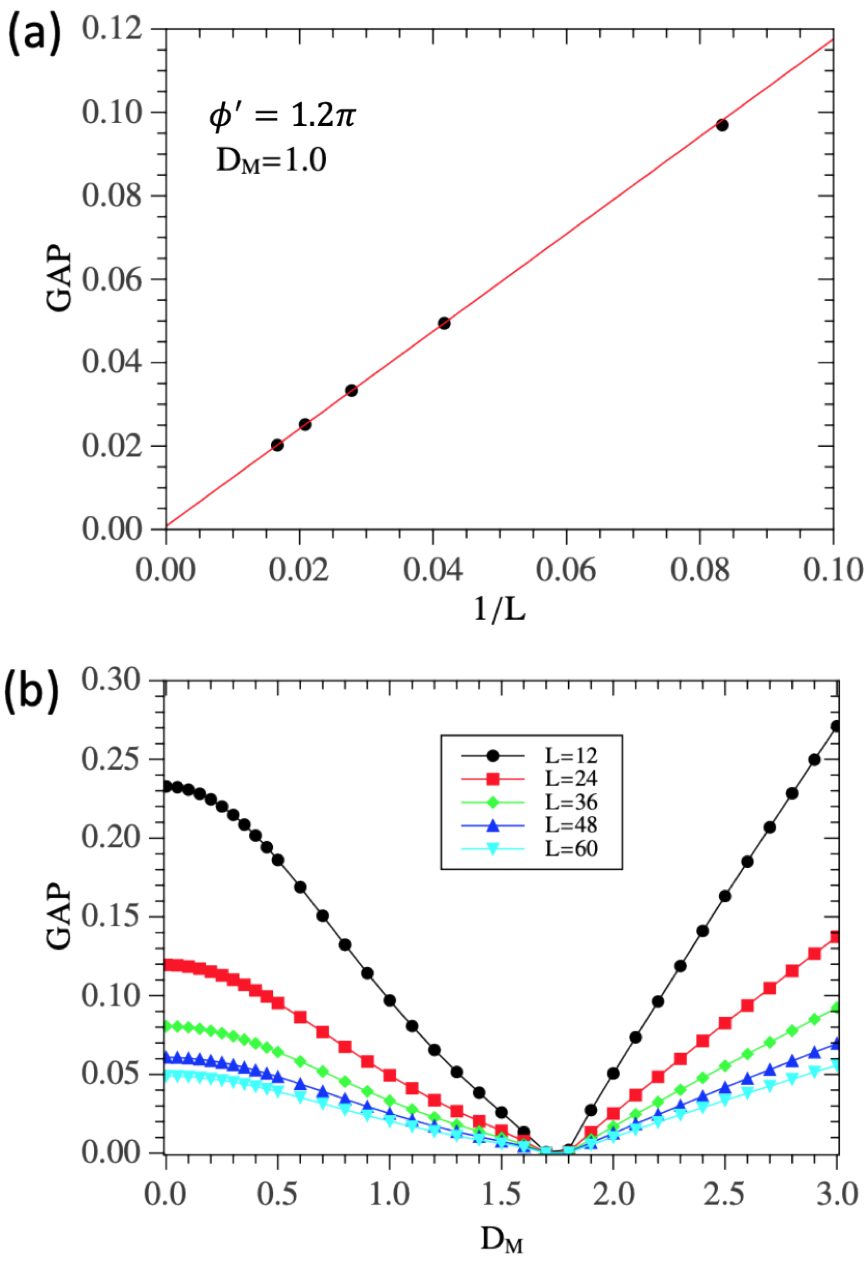}
\caption{
(a) Excitation gap as a function of   $1/L$  at  $\phi^\prime=1.2\pi$, $D_M=1.0$,
(b) gap vs. $D_M$ with fixed $\theta=1.2\pi$
for several system sizes marked with different colored lines. 
} 
\label{fig:DM_gap_scaling}
\end{center}
\end{figure}

\subsubsection{Scaling of the finite size gap}
\label{subsec:scaling_gap}

\begin{figure*}[ht]
\begin{center}
\includegraphics[width=18cm]{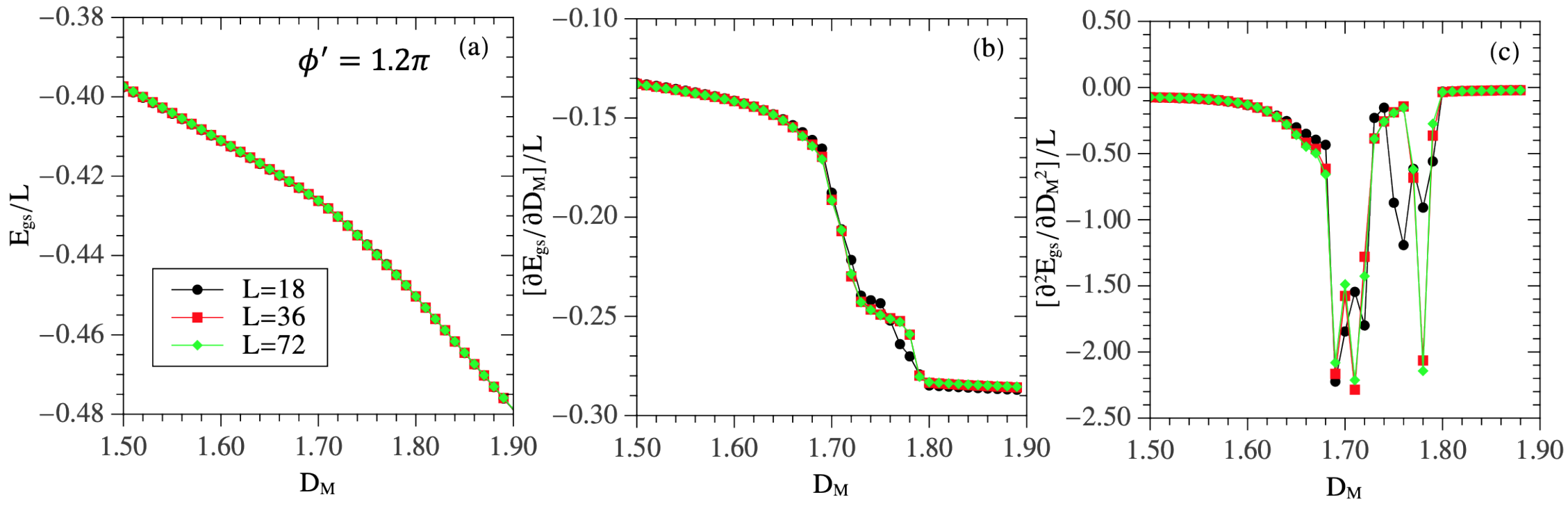}
\caption{(a) Ground state energy per site  $E_{gs}/L$, (b) first order derivative $\frac{1}{L}\frac{\partial E_{gs}}{\partial D_M}$,
(c) second order derivative $\frac{1}{L}\frac{\partial^2 E_{gs}}{\partial D_M^2}$ as functions of $D_M$ for $L=18$ (black), $L=36$ (red) and $L=72$ (green) at  $\phi^\prime=1.2\pi$. 
} 
\label{fig:DM_phase_transition}
\end{center}
\end{figure*}

We also investigate the scaling of the finite size gap,
which is predicted to be $\sim1/L$ according to CFT. 
Fig. \ref{fig:DM_gap_scaling} (a) shows the excitation gap as a function of $1/L$ at $\phi^\prime=1.2\pi$, $D_M=1.0$,
which is clearly very linear, consistent with the CFT prediction. 
Fig. \ref{fig:DM_gap_scaling} (b) shows the scaling of the gap by tuning $D_M$ where $\phi^\prime$ is fixed as $\phi^\prime=1.2\pi$.
As can be seen from Fig. \ref{fig:DM_gap_scaling} (b), the range of the emergent SU(2)$_1$ phase is rather large,
except the narrow region for $D_M\in [1.68,1.80]$ where the $1/L$ scaling fails.  
Therefore, the regions for $D_M\leq 1.68$ and $D_M\geq 1.80$ in Fig. \ref{fig:DM_gap_scaling} (b)  belong to the ``Emergent SU(2)$_1$ I" and ``Emergent SU(2)$_1$ II" phases in Fig. \ref{fig:phase_KGDM}, respectively.

\subsubsection{Phase transitions}
\label{subsubsec:phase_transit}

\begin{figure*}[ht]
\begin{center}
\includegraphics[width=13cm]{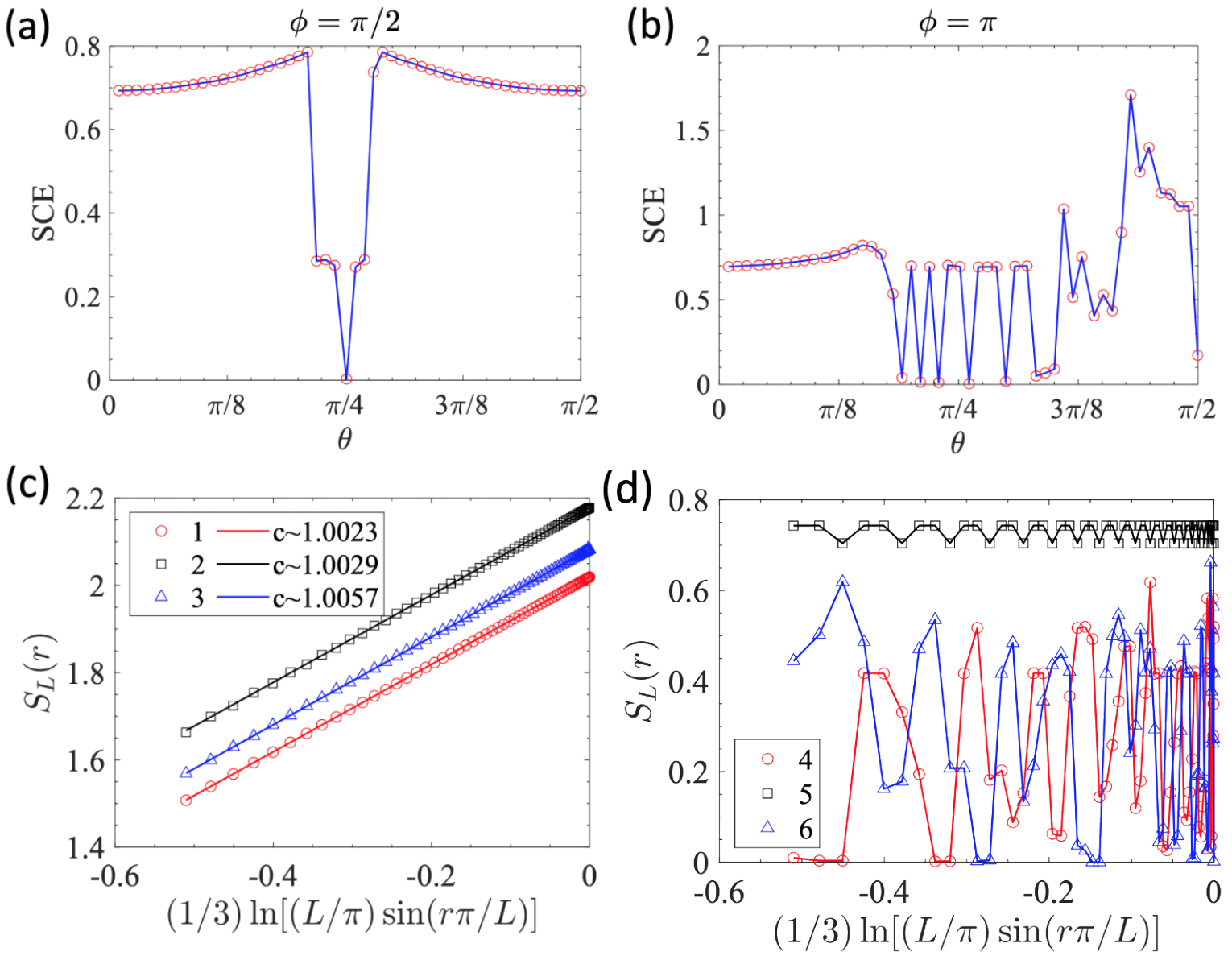}
\caption{SCE as a function of $\theta$ at (a) $\phi=\pi/2$ and (b) $\phi=\pi$, and plots of entanglement entropies as functions of $\frac{1}{3} \ln \left[ \frac{L}{\pi} \sin\left( \frac{\pi r}{L} \right) \right]$ for points (c) ``$1,2,3$" and (d) ``$4,5,6$" in Fig. \ref{fig:phase_KGDM}.
DMRG numerical simulations are performed on $L=36$ sites for (a,b) and $L=144$ sites for (c,d), using periodic boundary conditions. 
The bond dimension $m$ and truncation error $\epsilon$ are taken as $m=1200$ and $\epsilon=10^{-8}$.
} 
\label{fig:phase_boundary}
\end{center}
\end{figure*}

We take a closer look at the region $D_M\in [1.68,1.80]$ in Fig. \ref{fig:DM_gap_scaling} (b), where the system does not show a Luttinger liquid behavior. 
Fig. \ref{fig:DM_phase_transition} (a), (b), (c) show the ground state energy per site  $E_{gs}/L$, the first order derivative $\frac{1}{L}\frac{\partial E_{gs}}{\partial D_M}$,
and the second order derivative $\frac{1}{L}\frac{\partial^2 E_{gs}}{\partial D_M^2}$, respectively, as functions of $D_M$ at $\phi^\prime=1.2\pi$. 
As can be seen from Fig. \ref{fig:DM_phase_transition},
the ground state energy is non-analytic at $D_M=1.68$ and $1.80$, indicating phase transitions at those two points. 

\subsection{Phase boundaries}
\label{subsec:phase_transition}

We turn back to the $(\theta,\phi)$ parameterization in Eq. (\ref{eq:parametrize_KGDM}) in this subsection. 

\subsubsection{Single-copy entanglement}
\label{subsubsec:SCE}

A more sensitive probe for determining phase transitions is the single-copy entanglement (SCE) \cite{Eisert2005},
defined as 
\bea
\text{SCE}=-\ln \lambda_1,
\eea
where $\lambda_1$ is the largest eigenvalue of the reduced density matrix for half of the chain.
For instance, Fig. \ref{fig:phase_boundary} (a,b) shows SCE as a function of $\theta$ by fixing $\phi$ to be $0.5\pi$ and $\pi$, respectively,
obtained from DMRG calculations on systems of $L=36$ sites using periodic boundary conditions. 
The emergent SU(2)$_1$ phases correspond to regions in which SCE is a smooth function of $\theta$,
and the phase boundaries can be determined from the points where smoothness is lost. 
For example, when $\phi=0.5\pi$, as can be inspected from Fig. \ref{fig:phase_boundary} (a), 
all values of $\theta$ reside in the emergent SU(2)$_1$ phases except a small region surrounding $\theta=\pi/4$.
On the other hand, when $\phi=\pi$, only $\theta\in[0,0.17\pi)$ belongs to the emergent SU(2)$_1$ phase,
and other values of $\theta$ are outside of such phases. 
The red points in Fig. \ref{fig:phase_KGDM} are phase transition points determined from SCE in this way,
which gives the boundaries of the emergent SU(2)$_1$ phases as shown by the blue lines in the figure. 

To further confirm that the regions encircled by the blue lines in Fig. \ref{fig:phase_KGDM} are indeed gapless phases having emergent SU(2)$_1$ conformal symmetries at low energies, we have calculated the central charge values for three representative points within the emergent SU(2)$_1$ phases denoted as ``$1,2,3$" in Fig. \ref{fig:phase_KGDM}, whose $(\theta,\phi)$ coordinates are given by: $(\theta=0.25\pi, \phi=0.1\pi)$ for point 1, $(\theta=0.23\pi, \phi=0.8\pi)$ for point 2, and $(\theta=0.3\pi, \phi=1.2\pi)$ for point 3. As can be seen from Fig. \ref{fig:phase_boundary} (c), the central charges at these three points are all very close to $1$, consistent with the emergent SU(2)$_1$ predictions. 

On the other hand, as can be seen from Fig. \ref{fig:phase_boundary} (d), no value of central charge can be reliably extracted  for three representative points outside of the emergent SU(2)$_1$ phases marked by ``$4,5,6$" in Fig. \ref{fig:phase_KGDM}, whose $(\theta,\phi)$ coordinates are given by  $(\theta=0.25\pi, \phi=0.5\pi)$ for point 4, $(\theta=0.3\pi, \phi=\pi)$ for point 5, and  $(\theta=0.25\pi, \phi=1.5\pi)$ for point 6. 
Notice that points 5 and 6 are the exactly solvable points as discussed in Sec. \ref{subsec:solvable}, having an exponentially large ground state degeneracy. 
Therefore, it is expected that they do not exhibit Luttinger liquid behaviors. 

\subsubsection{Regions near $(\theta=\pi/4,\phi=\pi/2)$}
\label{subsubsec:non_SU2_pt4}

\begin{figure}
\begin{center}
\includegraphics[width=7cm]{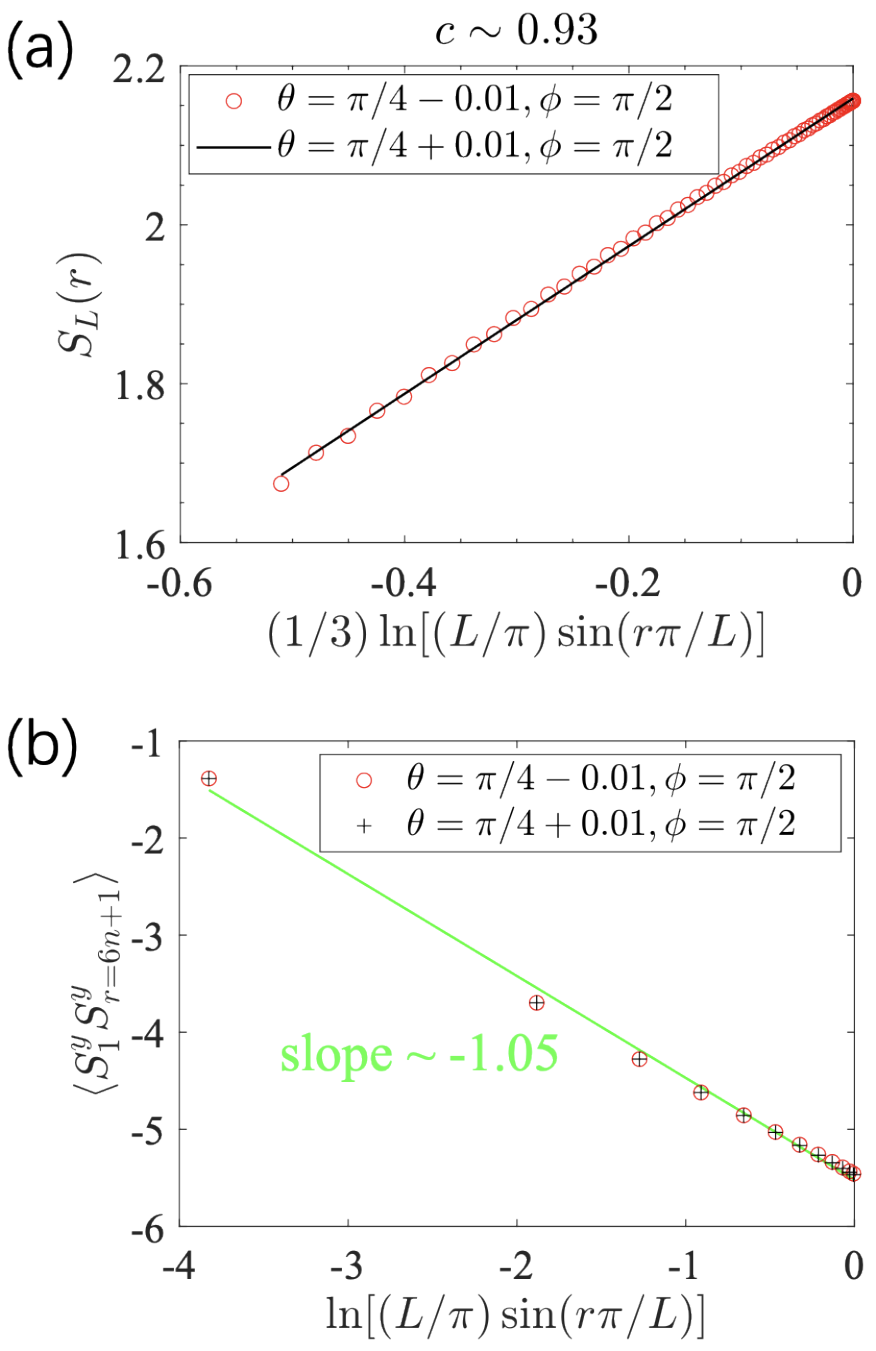}
\caption{(a) Entanglement entropy $S_L(x)$ vs.  $\frac{1}{3}\ln[\frac{L}{\pi}\sin(\frac{\pi r}{L})]$ and (b) spin correlation function $\langle S^y_1S^y_{r}\rangle$ ($r=1+6n$) as a function of  $\frac{L}{\pi}\sin(\frac{\pi r}{L})$ on a log-log scale,  at $\theta=0.25\pi\pm0.01, \phi=0.5\pi$.
DMRG numerics are performed on systems of $L=144$ sites with periodic boundary conditions. 
The bond dimension $m$ and truncation error $\epsilon$ are taken as $m=1200$ and $\epsilon=10^{-8}$.
} 
\label{fig:near_point_4}
\end{center}
\end{figure}

As can be seen from Fig. \ref{fig:phase_KGDM}, the narrow region between the two dashed lines around point 4 is not in the emergent  SU(2)$_1$ phase, which is based on DMRG numerics on SCE for systems of $L=36$ sites as discussed in Sec. \ref{subsubsec:SCE}.
To test whether the existence of such narrow region is a finite size artifact and to investigate the nature of the physics in this parameter region, we have increased the system size to $L=72$ sites and calculated the entanglement entropy $S_L(r)$ as well as spin correlation function $\langle S^y_1S^y_{r}\rangle$ at $(\theta=\pi/4\pm 0.01, \phi=\pi/2)$, using DMRG numerics under periodic boundary conditions.
The convergence turns out to be extremely slow, indicating numerical difficulties in this parameter region.  

The numerical results are shown in Fig. \ref{fig:near_point_4}, with central charge value $c=0.93$ and exponent $\nu$ of the correlation function $\langle S^y_1S^y_{r}\rangle$ as $\nu=1.05$. 
Notice that SU(2)$_1$ CFT predicts $c$ and $\nu$ to be $c=1$ and $\nu=1$ (up to logarithmic correction discussed in Sec. \ref{subsec:correlation}). 
Therefore, our numerics indicate that the range of the narrow non-SU(2)$_1$ phase surrounding point 4 shrinks as the system size increases,
which is the reason why the phase boundaries in Fig. \ref{fig:phase_KGDM} in this region are plotted as dashed lines. 
Whether such narrow phase persists in the thermodynamic limit remains to be further analytically explored.

\subsubsection{Regions near $\phi=\pi$}

\begin{figure*}[ht]
\begin{center}
\includegraphics[width=18cm]{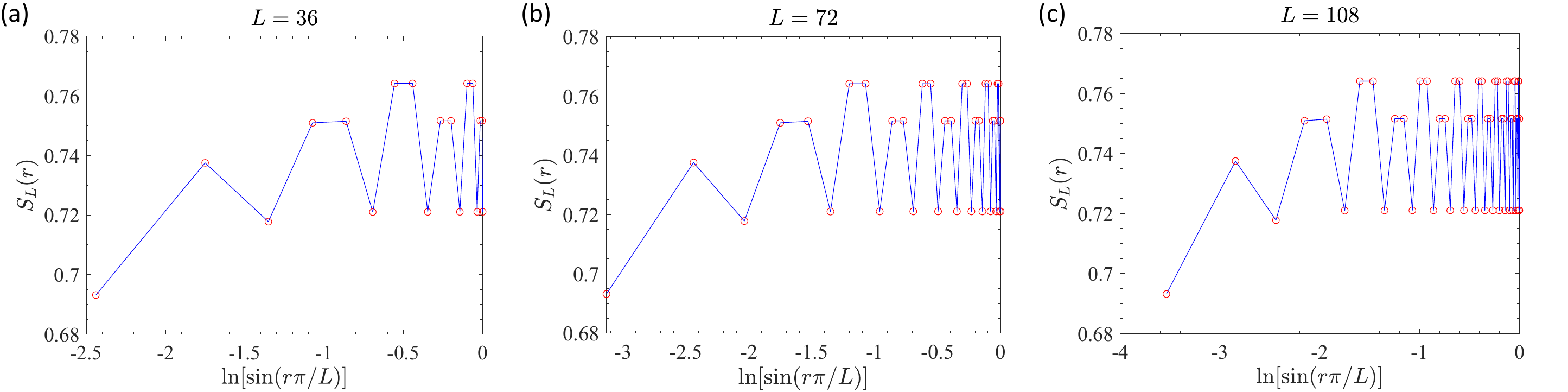}
\caption{Entanglement entropy $S_L(r)$ vs. $\ln [\sin(\frac{\pi r}{L})]$ at $(\theta=\pi/4\pm 0.01, \phi=\pi/2)$ with system sizes (a) $36$, (b) $72$, (c) $108$.
DMRG numerics are performed under periodic boundary conditions, with bond dimension $m$ and truncation error $\epsilon$ taken as $m=1200$ and $\epsilon=10^{-10}$.
} 
\label{fig:c_size_point5}
\end{center}
\end{figure*}

\begin{figure}
\begin{center}
\includegraphics[width=8cm]{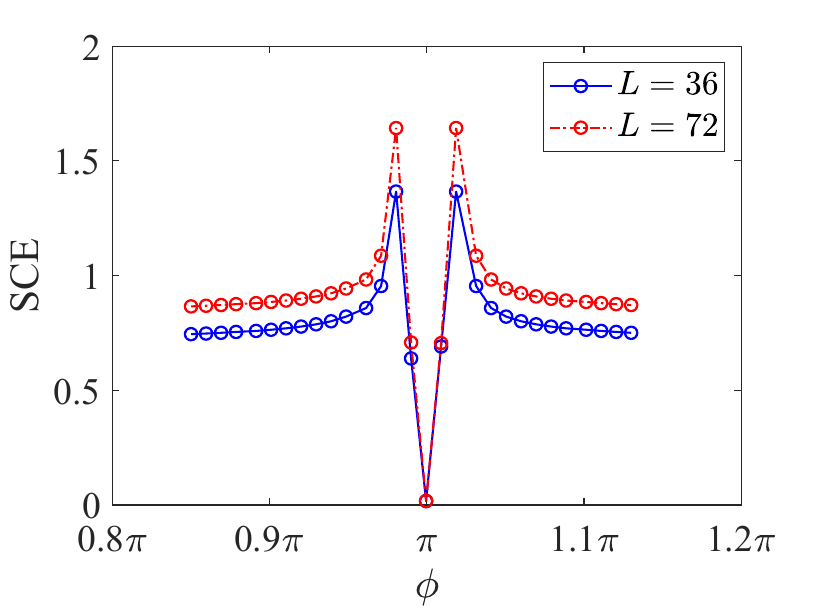}
\caption{SCE vs. $\phi$ at $\theta=0.3\pi$ around $\phi=\pi$ for system sizes $L=36,72$.
DMRG numerics are performed under periodic boundary conditions, with bond dimension $m$ and truncation error $\epsilon$ taken as $m=1200$ and $\epsilon=10^{-8}$.
} 
\label{fig:SCE_5_36_72}
\end{center}
\end{figure}

The shrinking of the narrow non-SU(2)$_1$ phase near $(\theta=0.25\pi,\phi=0.5\pi)$ discussed in Sec. \ref{subsubsec:non_SU2_pt4} poses the concern whether the narrow phase surrounding the $\phi=\pi$ line in Fig. \ref{fig:phase_KGDM} is similarly a finite-size artifact. 
As shown in Fig. \ref{fig:c_size_point5}, we have numerically calculated the entanglement entropy $S_L(r)$ by increasing system size at the point $(\theta=0.3\pi,\phi=\pi-0.01)$. 
As can be inspected from Fig. \ref{fig:c_size_point5}, no reliable values of the central charge can be extracted for all system sizes $L=36,72,108$, clearly indicating a non-Luttinger-liquid behavior. 
To study how the range of the non-SU(2)$_1$ phase changes with system sizes, SCE as a function of $\phi$ around $\phi=\pi$ have been calculated at $\theta=0.3\pi$ for system sizes $L=36,72$, as shown in Fig. \ref{fig:SCE_5_36_72}.
The region between the two cusps in Fig. \ref{fig:SCE_5_36_72} corresponds to the non-SU(2)$_1$ phase.
It is clear from Fig. \ref{fig:SCE_5_36_72} that the range of this non-SU(2)$_1$ phase remains stable when the system size increases.
Based on the numerical results in Fig. \ref{fig:c_size_point5} and  Fig. \ref{fig:SCE_5_36_72}, we expect that the non-SU(2)$_1$ phase close to the $\phi=\pi$ line persists in the thermodynamic limit.

\section{Summary}
\label{sec:summary}

In summary, we have studied the spin-1/2 Kitaev-Gamma chain with a bond-dependent Dzyaloshinskii-Moriya interaction.
Using a two-site periodic unitary transformation which maps the Gamma and Dzyaloshinskii-Moriya terms into one another, the phase diagram of a pure Kitaev chain with a Dzyaloshinskii-Moriya term is obtained from the known results of the Kitaev-Gamma model.
More importantly, we are able to analytically demonstrate that there exist extended gapless phases in the phase diagram of the spin-1/2 Kitaev-Gamma chain with a nonzero Dzyaloshinskii-Moriya interaction,
and that such gapless phases have an emergent SU(2)$_1$ conformal symmetry at low energies.
The analytical predictions are supported by  our large-scale density matrix renormalization group simulations.
Our work is useful for understanding the effects of Dzyaloshinskii-Moriya interactions and electric fields in one-dimensional and quasi-one-dimensional generalized Kitaev spin models.

\begin{acknowledgments}
W.Y. acknowledges the startup funding at Nankai University. 
W.Y. and I.A. acknowledge support from NSERC Discovery Grant 04033-2016.
C.X. is partially supported by Strategic Priority Research Program of CAS (No. XDB28000000).
A.N. acknowledges computational resources and services provided by Compute Canada and Advanced Research Computing at the University of
British Columbia. A.N. acknowledges support from the Max Planck-UBC-UTokyo Center for Quantum Materials and the Canada First Research Excellence Fund
(CFREF) Quantum Materials and Future Technologies Program of the Stewart Blusson Quantum Matter Institute (SBQMI).
\end{acknowledgments}
 
\appendix

\begin{widetext}
\section{Perturbation Hamiltonian for nonzero $D_M$}
\label{app:derivation}

In this appendix, we treat $D_M$ as a perturbation on the Kitaev-Gamma model  and use Eq. (\ref{eq:nonsymmorphic_bosonize}) to express Eq. (\ref{eq:DeltaH_D}) in terms of the SU(2)$_1$ low energy degrees of freedom.
To proceed on, we need  several operator product expansions (OPE) in the SU(2)$_1$ theory.
Using the following OPE \cite{DiFrancesco1997},
\begin{flalign}
&\vec{J}_L(w) g(z,\bar{z})= -\frac{1}{2} \frac{\vec{\sigma} g(z,\bar{z}) }{w-z} +:\vec{J}_Lg: (z,\bar{z})+O(|z-w|),\nn\\
&\vec{J}_R(w) g(z,\bar{z})= \frac{1}{2} \frac{ g(z,\bar{z})  \vec{\sigma}}{w-z} +:\vec{J}_Rg: (z,\bar{z})+O(|z-w|),
\label{eq:OPE_Jg}
\end{flalign}
we obtain ($\alpha=x,y,z$),
\begin{eqnarray}
iJ_L^\alpha(x) \text{tr} (g(x+d)\sigma^\alpha)
&=&\frac{1}{4\pi  d}\epsilon+:J_L^\alpha N^\alpha:,\nn\\
iJ_R^\alpha(x) \text{tr} (g(x+d)\sigma^\alpha)
&=&\frac{1}{4\pi  d}\epsilon+:J_R^\alpha N^\alpha:,\nn\\
i\text{tr} (g(x)\sigma^\alpha)J_L^\alpha(x+d)
&=&-\frac{1}{4\pi  d}\epsilon+:J_L^\alpha N^\alpha:,\nn\\
i\text{tr} (g(x)\sigma^\alpha)J_R^\alpha(x+d)
&=&-\frac{1}{4\pi  d}\epsilon+:J_R^\alpha N^\alpha:,
\label{eq:OPE_2}
\end{eqnarray}
in which
\bea
\epsilon&=&\text{tr}(g), \nn\\
N^\alpha&=&i \text{tr}(g\sigma^\alpha),
\label{eq:epsilon_N}
\eea
and the symbol $:AB:$ means the $O(d^0)$ term in the full OPE $A(x+d)B(x)$. 

Plugging Eq. (\ref{eq:nonsymmorphic_bosonize}) and Eq. (\ref{eq:OPE_2}) into $S_i^\alpha S_{i+1}^\alpha$, we obtain
\bea
S_i^\alpha S_{i+1}^\alpha&=& a^2 D_i^\alpha D_{i+1}^\alpha [J_L^\alpha(x)+J_R^\alpha(x)][J_L^\alpha(x+a)+J_R^\alpha(x+a)]-aC_i^\alpha C_{i+1}^\alpha N^\alpha(x) N^\alpha(x+a) \nn\\
&&+(-)^i a^{3/2} \big[
-\frac{1}{2\pi a}(D_i^\alpha C_{i+1}^\alpha+C_i^\alpha D_{i+1}^\alpha)\epsilon +(C_i^\alpha D_{i+1}^\alpha-D_i^\alpha C_{i+1}^\alpha) :(J_L^\alpha+J_R^\alpha)N^\alpha:
\big].
\label{eq:bosonize_SS}
\eea
We note that in Eq. (\ref{eq:bosonize_SS}), the second and third lines are full OPE,  
however, their exact expressions are not important for us.
Then plugging Eq. (\ref{eq:bosonize_SS}) into Eq. (\ref{eq:DeltaH_D}) and summing over the  three sites within a unit cell, we arrive at
\begin{flalign}
\Delta H^\prime = -D_M(C_1D_2-C_2D_1)\int dx :(\vec{J}_L+\vec{J}_R)\cdot\vec{N}:.
\label{eq:Delta_H_JN}
\end{flalign}
Here we note that we have essentially performed a first order perturbative treatment by directly projecting the Hamiltonian into the low energy space.

\section{Staggered second nearest neighbor coupling}
\label{app:2nd_quadrupole}

We make some further comments about the $(\vec{J}_L+\vec{J}_R)\cdot \vec{N}$ term.
As can be seen from Eq. (\ref{eq:Delta_H_JN}), the $(\vec{J}_L+\vec{J}_R)\cdot\vec{N}$ term vanishes at the hidden SU(2)$_1$ symmetric point where $D_1=D_2$, $C_1=C_2$. 
In fact, the appearance of $(\vec{J}_L+\vec{J}_R)\cdot\vec{N}$ is an unusual feature of the Kitaev-Gamma model arising from the nonsymmorphic symmetries. 
It can appear in the SU(2) symmetric AFM Heisenberg model only at second nearest neighbor level.
The nonabelian bosonization formulas of the spin operators for the SU(2) symmetric case (i.e., $D_1=D_2$, $C_1=C_2$) are
\begin{flalign}
\frac{1}{a} S_i^\alpha=J_L^\alpha+J_R^\alpha
+(-)^{i+1} \frac{c}{2\pi a} N^\alpha,
\label{eq:nonabel_H}
\end{flalign}
where $c$ is a constant.
Plugging Eq. (\ref{eq:nonabel_H}) and Eq. (\ref{eq:OPE_2}) into $S_i^\alpha S_{i+2}^\alpha$, we obtain
\bea
S_i^\alpha S_{i+2}^\alpha&=&a^2 [J_L^\alpha(x) J_L^\alpha(x+2a)+J_R^\alpha(x) J_R^\alpha(x+2a)+2J_L^\alpha(x) J_R^\alpha(x+2a)]+\frac{c^2}{4\pi^2}N^\alpha(x) N^\alpha(x+2a)\nn\\
&&+(-)^{i+1} \frac{ca}{\pi} (:J_L^\alpha N^\alpha:+:J_R^\alpha N^\alpha:).
\label{eq:OPE_SS_i}
\eea
Using Eq. (\ref{eq:OPE_SS_i}), it is clear that
\begin{flalign}
(-)^{i+1}S^\alpha_i  S^\alpha_{i+2}+(-)^{i+2} S^\alpha_{i+1} S^\alpha_{i+3}=\frac{2ca}{\pi} :(J^\alpha_L+J^\alpha_R) N^\alpha:,
\label{eq:staggered_2nd_quadrupole}
\end{flalign}
which shows that $:(J^\alpha_L+J^\alpha_R) N^\alpha:$ corresponds to a staggered second nearest neighbor coupling  in the AFM Heisenberg model.

\section{$(\vec{J}_L+\vec{J}_R)\cdot \vec{N}$ as a total derivative}
\label{app:total_derivative}

In this appendix, we give a formal proof that  $:(J^\alpha_L+J^\alpha_R) N^\alpha:$ is a total derivative within the SU(2)$_1$ low energy theory. 
Since the Hamiltonian in the holomorphic and anti-holomorphic sectors are of the Sugawara form which generates the translations, we have
\bea
[\frac{1}{3} \int dx :\vec{J}_L\cdot \vec{J}_L:,g(w,\bar{w})]&=&\partial_w g(w,\bar{w}),
\label{eq:commutation_L}
\eea
and
\bea
[\frac{1}{3} \int dx :\vec{J}_R\cdot \vec{J}_R:,g(w,\bar{w})] &=&\partial_{\bar{w}} g(z,\bar{w}).
\eea
In what follows, we focus on the holomorphic sector, as the treatment for the anti-holomorphic sector is similar. 
For simplification of notation, we write the normal order product $:AB:$ as $(AB)$ in accordance with the notation in Ref. \onlinecite{DiFrancesco1997}.

The commutator in Eq. (\ref{eq:commutation_L}) can be deformed to a small loop centering around the point $w$ as
\bea
\partial_w g(w,\bar{w})
&=&\frac{1}{3}\sum_\alpha\oint_w dy (J_{L}^{ \alpha} J_{L}^{\alpha})(y) g(w,\bar{w}).
\label{eq:partial_g_w}
\eea
To evaluate Eq. (\ref{eq:partial_g_w}), we need the $\frac{1}{y-w}$ term in the OPE $(J_L^{\alpha} J_L^{\alpha})(y) g(w,\bar{w})$.
Define $\contraction{}{A}{(z)}{B}
A(z)B(w)$ to be  the singular parts of the OPE $A(z)B(w)$, i.e., the terms with coefficients $(y-w)^{-n}$ where $n\geq 1$.
According to the generalized Wick's theorem in  Ref. \onlinecite{DiFrancesco1997}, the singular terms in the OPE $(J_L^{\alpha} J_L^{\alpha})(y) g(w,\bar{w})$  can be evaluated as 
\bea
\contraction{(J_L^{\alpha}}{J}{{}_L^{\alpha})(y)}{g}
\contraction{(J_L^{\alpha} J_L^{\alpha})(y) g(w,\bar{w})=
\frac{1}{2\pi i}\oint_y \frac{dz}{z-y}
\big[J_L^{\alpha}(z)}{J}{{}_L^{\alpha}(y)}{g}
\contraction{(J_L^{\alpha} J_L^{\alpha})(y) g(w,\bar{w})=
\frac{1}{2\pi i}\oint_y \frac{dz}{z-y}
\big[J_L^{\alpha}(z)J_L^{\alpha}(y)g(w,\bar{w})+}{J}{{}_L^{\alpha}(z)}{g}
(J_L^{\alpha} J_L^{\alpha})(y) g(w,\bar{w})=
\frac{1}{2\pi i}\oint_y \frac{dz}{z-y}
\big[J_L^{\alpha}(z)J_L^{\alpha}(y)g(w,\bar{w})
+J_L^{\alpha}(z)g(w,\bar{w})J_L^{\alpha}(y)
\big].
\label{eq:singular_JJ_g}
\eea
Using Eq. (\ref{eq:OPE_Jg}), Eq. (\ref{eq:singular_JJ_g}) becomes
\begin{flalign}
\frac{1}{2\pi i}\oint_y \frac{dz}{z-y}
(-\frac{1}{2})\big[\frac{\sigma^\alpha J_L^{\alpha}(z) g(w,\bar{w})}{y-w}
+\frac{\sigma^\alpha g(w,\bar{w})J_L^{\alpha}(y)}{z-w}
\big].
\end{flalign}
Further using  Eq. (\ref{eq:OPE_Jg}), we obtain
\bea
\frac{g(w,\bar{w})}{2(y-w)^2}-\frac{\sigma^\alpha (J^{\alpha}_L g)(w,\bar{w})}{y-w}.
\label{eq:obtain}
\eea
Notice that only the $1/(y-w)$ term in Eq. (\ref{eq:obtain}) contributes to Eq. (\ref{eq:partial_g_w}),
leading to
\bea
\partial_w g(w,\bar{w})=-\frac{1}{3}\sum_\alpha \sigma^\alpha (J^{\alpha}_L g)(w,\bar{w}).
\label{eq:simplify_partial_g}
\eea
Taking the trace of Eq. (\ref{eq:simplify_partial_g}), we arrive at
\bea
i\partial_w \epsilon(w,\bar{w})=-\frac{1}{3}(\vec{J}_L\cdot \vec{N})(w,\bar{w}).
\eea

Similar calculations in the anti-holomorphic sector gives
\bea
i\partial_{\bar{w}} \epsilon(w,\bar{w})=\frac{1}{3}(\vec{J}_R \cdot \vec{N})(w,\bar{w}).
\eea
Hence
\bea
((\vec{J}_L+\vec{J}_R)\cdot \vec{N})(w,\bar{w})=-3i(\partial_w-\partial_{\bar{w}}) \epsilon(w,\bar{w}).
\eea
Notice that $\partial_z-\partial_{\bar{z}}=-i\partial_x$, 
therefore
\bea
\int dx ((\vec{J}_L+\vec{J}_R)\cdot \vec{N})=-3 \int dx \partial_x \epsilon,
\eea
which clearly vanishes.

\section{Symmetry transformations of the WZW fields}
\label{app:transform_WZW}

As discussed in Ref. \onlinecite{Yang2020},
the symmetry  transformation properties of the WZW fields $g$ and $\vec{J_L},\vec{J}_R$ under time reversal $T$, spatial translation $T_a$, inversion $I$ and global spin rotation $R$ ($\in SO(3)$) are given by
\begin{eqnarray}
T: &\epsilon(x)\rightarrow \epsilon(x), &\vec{N}(x)\rightarrow -\vec{N}(x),\nn\\
&\vec{J}_L(x)\rightarrow -\vec{J}_R(x), &\vec{J}_R(x)\rightarrow -\vec{J}_L(x), 
\label{eq:transformT}
\end{eqnarray}
\begin{eqnarray}
T_a: &\epsilon(x)\rightarrow -\epsilon(x), &\vec{N}(x)\rightarrow -\vec{N}(x),\nn\\
&\vec{J}_L(x)\rightarrow \vec{J}_L(x), &\vec{J}_R(x)\rightarrow \vec{J}_R(x), 
\label{eq:transformTa}
\end{eqnarray}
\begin{eqnarray}
I: & \epsilon(x)\rightarrow -\epsilon(-x), &\vec{N}(x)\rightarrow \vec{N}(-x),\nn\\
&\vec{J}_L(x)\rightarrow \vec{J}_R(-x), &\vec{J}_R(x)\rightarrow \vec{J}_L(-x), 
\label{eq:transformI}
\end{eqnarray}
\begin{eqnarray}
R: &\epsilon(x)\rightarrow \epsilon(x), &N^\alpha(x)\rightarrow R^{\alpha}_{\,\,\beta}N^\beta(x),\nn\\
&J^\alpha_L(x)\rightarrow R^{\alpha}_{\,\,\beta} J^\beta_L(x), &J^\alpha_R(x)\rightarrow R^{\alpha}_{\,\,\beta}J^\beta_R(x).
\label{eq:transformR}
\end{eqnarray}
in which $x$ is the spatial coordinate, and $R^{\alpha}_{\,\,\beta}$ ($\alpha,\beta=x,y,z$) is the matrix element of the $3\times 3$ rotation matrix $R$ at position $(\alpha,\beta)$. 

\section{Nonsymmorphic nonabelian bosonization formulas}
\label{app:bosonization}

The explicit expressions of the nonsymmorphic nonabelian bosonization formulas are
\bea
S_{1+6n}^x&=&D_2 (J_L^x+J_R^x)-C_2 N^x,\nn\\
S_{1+6n}^y&=&D_2 (J_L^y+J_R^y)-C_2 N^y,\nn\\
S_{1+6n}^z&=&D_1 (J_L^z+J_R^z)-C_1 N^z,
\label{eq:bosonize_1}
\eea
\bea
S_{2+6n}^x&=&D_2^\prime (J_L^x+J_R^x)+C_2^\prime N^x,\nn\\
S_{2+6n}^y&=&D_1^\prime (J_L^y+J_R^y)+C_1^\prime N^y,\nn\\
S_{2+6n}^z&=&D_2^\prime (J_L^z+J_R^z)+C_2^\prime N^z,
\label{eq:bosonize_2}
\eea
\bea
S_{3+6n}^x&=&D_1 (J_L^x+J_R^x)-C_1 N^x,\nn\\
S_{3+6n}^y&=&D_2 (J_L^y+J_R^y)-C_2 N^y,\nn\\
S_{3+6n}^z&=&D_2 (J_L^z+J_R^z)-C_2 N^z,
\label{eq:bosonize_3}
\eea
\bea
S_{4+6n}^x&=&D_2^\prime (J_L^x+J_R^x)+C_2^\prime N^x,\nn\\
S_{4+6n}^y&=&D_2^\prime (J_L^y+J_R^y)+C_2^\prime N^y,\nn\\
S_{4+6n}^z&=&D_1^\prime (J_L^z+J_R^z)+C_1^\prime N^z,
\label{eq:bosonize_4}
\eea
\bea
S_{5+6n}^x&=&D_2 (J_L^x+J_R^x)-C_2 N^x,\nn\\
S_{5+6n}^y&=&D_1 (J_L^y+J_R^y)-C_1 N^y,\nn\\
S_{5+6n}^z&=&D_2 (J_L^z+J_R^z)-C_2 N^z,
\label{eq:bosonize_5}
\eea
\bea
S_{6+6n}^x&=&D_1^\prime (J_L^x+J_R^x)+C_1^\prime N^x,\nn\\
S_{6+6n}^y&=&D_2^\prime (J_L^y+J_R^y)+C_2^\prime N^y,\nn\\
S_{6+6n}^z&=&D_2^\prime (J_L^z+J_R^z)+C_2^\prime N^z,
\label{eq:bosonize_6}
\eea
in which $J^\alpha_L$, $J_R^\alpha$ and $N^\alpha$ are defined in Eq. (\ref{eq:J_LR}) and Eq. (\ref{eq:epsilon_N}).

\end{widetext}



\end{document}